%% file: main.tex
\documentclass[11pt]{article}

\usepackage[a4paper,margin=1in]{geometry}
\usepackage{microtype}
\usepackage{parskip}
\usepackage{placeins}

\usepackage{siunitx}
\usepackage{amsmath,amssymb,amsfonts}
\usepackage{algorithm}
\usepackage{algpseudocode}
\usepackage{booktabs}
\usepackage{graphicx}
\usepackage{subcaption}
\usepackage{hyperref}
\usepackage{xcolor}
\usepackage{multirow}
\usepackage{url}
\usepackage{cite}
\usepackage{bm}
\usepackage{authblk}

\newtheorem{lemma}{Lemma}

\title{Gradient Inversion through Linear and Affine Subspace Constraints}
\author[]{Konstantinos Bouros\thanks{Correspondence: \texttt{kbouros@tuc.gr}}}
\author[]{ Athanasios P. Liavas}
\affil[]{School of Electrical and Computer Engineering \authorcr
Technical University of Crete}
\date{}

\begin{document}

\maketitle

\input{section_files/abstract}
\input{section_files/1.introduction}
\input{section_files/2.related_work}
\input{section_files/3.method_fc}
\input{section_files/4.method_cnn}
\input{section_files/5.experiments}
\input{section_files/6.limitations_future_work}
\input{section_files/7.conclusion}

\section*{Declaration of AI-assisted tools}
ChatGPT was used to assist with parts of the code implementation and debugging, and to improve the wording and clarity of the manuscript. The authors take full responsibility for the implementation, experimental results, and manuscript content.

% References
\FloatBarrier
\bibliographystyle{IEEEtran}
\bibliography{references}
\FloatBarrier

% Appendix
\input{section_files/appendix}

\end{document}

%% file: section_files/abstract.tex
\begin{abstract}
Federated learning enables multiple clients to collaboratively train a machine-learning model without directly sharing their local datasets. However, as prior works have shown, the gradients exchanged during training may reveal sensitive information about client data. Most existing attacks usually optimize dummy inputs to match the true client gradients (Gradient Inversion Attacks), while some attempt to reconstruct them based on the observation that aggregate gradients from a fully connected (FC) layer are a linear combination of its inputs. In this work, we extend this observation by showing that aggregate gradients from an FC layer with the correct scaling are actually an affine combination of its inputs. We evaluate a gradient inversion attack that leverages either known linear-subspace information or its affine-subspace counterpart at each FC layer to improve reconstruction accuracy. Our results show that subspace-based methods can substantially improve reconstruction quality over the chosen baselines in certain settings. We further demonstrate that moderate overestimation of the batch size can generally improve or preserve reconstruction quality, even when the true batch size is unknown. We additionally examine reconstruction under gradient perturbations introduced by a differential privacy (DP) mechanism, providing insight into the effectiveness of the considered formulations under noisy conditions.
\end{abstract}

%% file: section_files/1.introduction.tex
\section{Introduction}
Federated learning allows a set of clients to collaboratively optimize a shared model while keeping their local training data private. In particular, this is achieved by performing the training process locally on clients and sending the shared gradients to a central server for aggregation. However, although this training paradigm avoids direct data sharing, it does not automatically guarantee privacy. Gradients themselves may contain sensitive information about the samples used to compute them. 

Zhu et al.~\cite{dlg} first demonstrated that private data can be reconstructed from the corresponding gradients by exploiting this vulnerability. This technique and related methods are now known as Gradient Inversion Attacks (GIAs). The reconstruction accuracy of these attacks tends to be sensitive to the input and batch size, and additional information or a reformulation of the problem is usually required to improve the performance of the attack.

Cocktail Party Attack~\cite{cpa} is one such approach, which frames gradient inversion as a blind source separation (BSS) problem by leveraging the crucial insight that aggregate gradients from a fully connected (FC) layer are a linear combination of its inputs. This method achieves significant improvement of accuracy but essentially relies on extracted information from a single FC layer.

In an attempt to further improve private data reconstruction performance, we propose an algorithm that follows the GIA paradigm and utilizes geometric information from each FC layer in the architecture. In particular, we make the observation that each row of the aggregated weight gradients from an FC layer, divided by the corresponding aggregated bias gradient entry, is an affine combination of the layer inputs. Our algorithm preserves the optimization framework and objective of Geiping et al. \cite{ig}, while incorporating gradient-induced linear or affine subspaces for each intermediate output as regularizers.

Lastly, although prior work reasonably assumes that the attacker knows the batch size, exact knowledge may not be necessary. In the noiseless setting, the batch size can typically be inferred directly from the rank of an FC layer's weight gradient matrix due to the aforementioned gradient structure under suitable full-rank conditions. However, exact batch-size information may not always be available to the attacker, particularly when the observed gradients are perturbed.
We therefore also evaluate the sensitivity of the attacks to inaccurate batch-size information. Our results show that, within the range of batch-size overestimation considered, reconstruction can remain successful and may occasionally improve. Note that this improvement is measured by retaining only the best $B$ reconstructions, where $B$ is the true batch size, and averaging the evaluation metrics over this subset. This assumes that the attacker can identify and discard poor reconstructions, which is typically straightforward in image-based settings.

Our main contributions include:
\begin{itemize}
    \item The observation that each row of the aggregated weight gradient of an FC layer, divided by the corresponding nonzero bias-gradient entry, is an affine combination of the layer inputs.
    
    \item Differentiable input and embedding subspace regularizers that exploit the linear or affine structure of FC-layer gradients. These regularizers can potentially be incorporated into optimization-based GIAs targeting architectures that contain FC layers.
    
    \item An implementation and evaluation of a gradient inversion attack that exploits these constraints across multiple FC layers, considering different architectures, datasets, and gradient perturbation levels. Our implementation builds upon the CPA codebase \cite{cpa} and is available at \url{https://github.com/kbouros/subspace-gradient-inversion}.
    
    \item An experimental analysis of how exact and potentially inaccurate batch-size information available to the attacker affects reconstruction performance.
\end{itemize}
In this work, we focus on the honest-but-curious setting where in-distribution data are not available to the adversary.

%% file: section_files/2.related_work.tex
\section{Related Work}
In this section, we present an overview of the most relevant prior gradient inversion attacks (GIAs) and low-rank inversion techniques.

\subsection{Gradient Matching}
The first gradient inversion attack was introduced by Zhu et al.~\cite{dlg}. They demonstrated that sensitive training data can be reconstructed by optimizing an input-label pair $(\bm{x}, \bm{y})$ whose gradients closely match the observed gradients.
The DLG optimization problem is formulated as
\begin{equation}
\begin{gathered}
\min_{\bm{x},\bm{y}} \;
 D_{\mathrm{GM}}(\bm{x},\bm{y}), \\
D_{\mathrm{GM}}(\bm{x},\bm{y})
:=
d\!\left(
\nabla_{\bm{\theta}}
\mathcal{L}_{\bm{\theta}}(\bm{x},\bm{y}),
\nabla_{\bm{\theta}}
\mathcal{L}_{\bm{\theta}}^{\mathrm{true}}
\right),
\end{gathered}
\end{equation}
where $d$ denotes a distance metric between the candidate and observed gradient vectors. Zhu et al.~\cite{dlg} use the $\ell_2$ distance for this purpose. Geiping et al.~\cite{ig} later proposed replacing the $\ell_2$ distance with cosine distance, making the optimization insensitive to the gradients' magnitudes. They also incorporated total variation (TV) regularization, which promotes the spatial smoothness commonly observed in natural images.

\subsection{Generative Models and Batch Norm Statistics}
Several subsequent works have sought to improve reconstruction quality, particularly for higher-resolution images and larger batches, by incorporating stronger priors into the inversion objective. For example, GradInversion uses the running statistics of batch-normalization (BN) layers as a distributional prior~\cite{gradinversion}, whereas GIAS constrains the reconstruction using a pretrained generative model~\cite{gias}. However, BN-statistics-based priors are applicable only to architectures that contain BN layers. In practical federated learning settings, the shared model may not include such layers, either for privacy considerations or because they are unsuitable for the task or architecture.

Generative-model-based approaches face a similar practical limitation. They typically require access to a dataset whose distribution is the same or sufficiently close to that of the private training data in order to train the generative model. This information may not be available to an attacker, limiting the applicability of these attacks under substantial distribution shift.

More recent methods relax these assumptions in different ways. GIFD~\cite{gifd} improves the out-of-distribution (OOD) flexibility of GAN-based inversion by optimizing intermediate generator features, but it still requires a pretrained GAN that provides a useful prior for the target domain. GI-PIP~\cite{gi-pip} instead uses an anomaly-detection prior trained on a small auxiliary dataset and is more tolerant to distribution mismatch. However, both methods still depend on auxiliary information and therefore relax, rather than eliminate, the distributional assumptions.

\subsection{Low-Rank Gradient Inversion}
Other methods exploit low-rank structure in fully connected or self-attention gradients to reconstruct private inputs approximately~\cite{cpa,tiger} or, under suitable assumptions, exactly~\cite{spear,spear++,dager}. CPA~\cite{cpa} models the gradient of a fully connected layer as a linear mixture of its inputs and reformulates inversion as a blind source separation problem.

SPEAR~\cite{spear} and SPEAR++~\cite{spear++} show that exact batch reconstruction from fully connected-layer gradients is possible under suitable rank, dimensionality, and sparsity conditions. Both exploit the sparsity induced by ReLU activations to identify the correct low-rank factorization, while SPEAR++ improves the scalability of the original approach. Their applicability, however, remains limited to settings with sufficiently sparse activations and favorable layer dimensions, and the exact-reconstruction guarantees do not directly extend to non-sparse activation functions.

DAGER~\cite{dager} extends low-rank gradient inversion to transformer language models by exploiting information from selected self-attention gradients together with the discrete structure of token embeddings. Earlier low-rank attacks typically rely on effectively extracting information from a single or limited set of layers. TIGER~\cite{tiger} builds upon the low-rank observation and combines subspace information from multiple transformer layers, improving reconstruction quality and robustness to DP noise.

\subsection{Malicious Server Setting}
In the malicious-server setting, the server can deviate from the prescribed federated-learning protocol by modifying the model architecture or parameters to isolate, amplify, or directly expose individual training samples~\cite{robbing_the_fed,curious_abandon_honesty,fishing_for_user_data}. VGIA~\cite{vgia} similarly manipulates model parameters and explicitly exploits the linear structure of FC-layer gradients together with ReLU-induced geometric partitioning to achieve verifiable isolation and reconstruct individual records.

Although these attacks can achieve significantly more effective extraction than passive gradient inversion, they require an active adversary and may be exposed if clients inspect or audit the model they receive. Since this work focuses on the weaker honest-but-curious setting, we do not examine malicious-server attacks further.

\subsection{Prior Work Limitations}
Most prior gradient-inversion attacks do not exploit the linear structure of FC-layer gradients. Moreover, although the underlying gradient equations are well established, the affine-subspace constraints they imply have, to our knowledge, not been explicitly recognized or leveraged in gradient inversion. Among methods that do exploit this structure, SPEAR and SPEAR++~\cite{spear,spear++} provide powerful exact-reconstruction attacks under suitable conditions. These methods, however, rely on ReLU-induced sparsity as a key source of information. While ReLU remains widely used, more general threats directly applicable to architectures with non-sparse activations should also be considered.
 
Other low-rank attacks, such as CPA~\cite{cpa}, exploit information from a selected fully connected layer but do not systematically combine constraints from multiple layers. The method closest to our approach is the recent TIGER~\cite{tiger}, which uses linear-subspace information from multiple transformer layers but is specialized to the text domain. Unlike TIGER, which replaces full-gradient matching with a subspace-based objective, our approach retains the gradient-matching objective and incorporates linear or affine subspace information as additional regularization in the image setting. More generally, the proposed subspace regularizers may be combined with other existing or future optimization-based inversion methods.

%% file: section_files/3.method_fc.tex
\section{Subspace-Constrained Gradient Inversion}
In this section, we analyze the affine structure of aggregate gradients in FC layers and describe the algorithm in detail for FC networks. In Section \ref{section 4}, we explain how the algorithm can be adapted to convolutional neural networks (CNNs) using an approach similar to that of CPA~\cite{cpa}.

\subsection{Affine Structure of FC Layers}
Consider a neural network with $L$ FC layers. Let $\bm{x}^{(i)} \in \mathbb{R}^{d_0}$
denote the $i$-th input in a batch of size $B$ and define
$\bm{h}^{(i)}_0 := \bm{x}^{(i)}.$
For an arbitrary layer $l \in \{1,\dots,L\}$, let
\[
\bm{a}^{(i)}_l = \bm{W}_l \bm{h}^{(i)}_{l-1} + \bm{b}_l,
\qquad
\bm{h}^{(i)}_l = \phi_l\!\left(\bm{a}^{(i)}_l\right),
\]
where $\bm{W}_l \in \mathbb{R}^{d_l \times d_{l-1}}$ and
$\bm{b}_l \in \mathbb{R}^{d_l}$ are the weights and biases of layer
$l$, respectively. Let the total mean loss over the batch be
\[
\mathcal{L}
=
\frac{1}{B}\sum_{i=1}^{B} \ell^{(i)} .
\]
\textbf{Layer-$l$ gradient identities.}
Fix a layer $l \in \{1,\dots,L\}$. Define the backpropagated signal
for sample $i$ at layer $l$ as
\[
\bm{\delta}^{(i)}_l := \frac{\partial \mathcal{L}}{\partial \bm{a}^{(i)}_l} =
\frac{1}{B} \frac{\partial \ell^{(i)}}{\partial \bm{a}^{(i)}_l}
\in \mathbb{R}^{d_l}. 
\]
By the chain rule, the per-sample gradients of the layer-$l$ parameters with respect to $\ell^{(i)}$ satisfy
\[
\nabla_{\bm{W}_l} \ell^{(i)}=\frac{\partial \ell^{(i)}}{\partial \bm{a}^{(i)}_l}{\bm{h}^{(i)}_{l-1}}^\top = B \cdot \bm{\delta}^{(i)}_l {\bm{h}^{(i)}_{l-1}}^\top, \quad 
\nabla_{\bm{b}_l} \ell^{(i)} = \frac{\partial \ell^{(i)}}{\partial \bm{b}_l} =
\frac{\partial \ell^{(i)}}{\partial \bm{a}^{(i)}_l} = B \cdot \bm{\delta}^{(i)}_l .
\]
Because the total loss is the mean over the batch, the observed gradients are
\[
\bm{G}_{\bm{W}_l} := \nabla_{\bm{W}_l}\mathcal{L} = \frac{1}{B}\sum_{i=1}^B \nabla_{\bm{W}_l} \ell^{(i)} = \sum_{i=1}^B \bm{\delta}^{(i)}_l {\bm{h}^{(i)}_{l-1}}^\top,
\quad
\bm{g}_{\bm{b}_l} := \nabla_{\bm{b}_l}\mathcal{L} = \frac{1}{B}\sum_{i=1}^B \nabla_{\bm{b}_l} \ell^{(i)} = \sum_{i=1}^B \bm{\delta}^{(i)}_l .
\]
For each neuron $j \in \{1,\dots,d_l\}$, denote the $j$-th row of $\bm{G}_{\bm{W}_l}$ and the corresponding bias-gradient entry, respectively, by
\[
\bm{r}^{\top}_{j} := \left(\bm{G}_{\bm{W}_l}\right)_{j,:} \in \mathbb{R}^{d_{l-1}}, \quad
s_{j} := \left(\bm{g}_{\bm{b}_l}\right)_j .
\]
Then, the gradient identities become
\[
\bm{r}^{\top}_{j} = \sum_{i=1}^{B} \delta^{(i)}_{l,j} 
{\bm{h}^{(i)}_{l-1}}^\top, \qquad s_{j} = \sum_{i=1}^{B} \delta^{(i)}_{l,j}.
\]
\textbf{Normalized gradients.}
Assume $s_{j} \neq 0$. Define the normalized vector
$ \bm{u}_{j} := \frac{\bm{r}_{j}}{s_{j}}$. 
Substituting the expressions above yields
\[
\bm{u}_{j} = \frac{ \sum_{i=1}^{B}\delta^{(i)}_{l,j} \bm{h}^{(i)}_{l-1}} {\sum_{i=1}^{B}\delta^{(i)}_{l,j}} = 
\sum_{i=1}^{B} \frac{ \delta^{(i)}_{l,j} } {\sum_{k=1}^{B}\delta^{(k)}_{l,j}} \; \bm{h}^{(i)}_{l-1}.
\]
Define the coefficients
\[
\alpha_{j,i} := \frac{ \delta^{(i)}_{l,j} }{ \sum_{k=1}^{B} \delta^{(k)}_{l,j}}.
\]
Then,
\[
\sum_{i=1}^{B} \alpha_{j,i} = 1
\quad \text{and} \quad
\bm{u}_j
=
\sum_{i=1}^{B}
\alpha_{j,i} \bm{h}^{(i)}_{l-1}.
\]

\begin{lemma}[Affine-hull characterization]
Let
\[
\mathcal{J} = \{j \in \{1,\dots,d_l\} \mid s_j \neq 0\}.
\]
For every $j \in \mathcal{J}$, $\bm{u}_j$ is an
affine combination of the layer-$l$ inputs, and hence
\[
\{\bm{u}_j \mid j \in \mathcal{J}\}
\subseteq
\mathrm{aff}\{\bm{h}_{l-1}^{(1)},\dots,\bm{h}_{l-1}^{(B)}\}.
\]
\end{lemma}

\textbf{Remark.} For the first layer specifically, where $\bm{h}^{(i)}_0 = \bm{x}^{(i)}$, each normalized gradient vector $\bm{u}_j$ belongs to the affine hull of the original inputs:
\[\{\bm{u}_j \mid j \in \mathcal{J}\}\subseteq
\mathrm{aff}\{\bm{x}^{(1)},\dots, \bm{x}^{(B)}\}.\]

\textbf{Consequence.}
The normalized layer-$l$ gradient rows $\{\bm{u}_j \mid j \in \mathcal{J}\}$ lie in the affine hull of the batch layer inputs. Therefore, under suitable rank conditions and provided that $|\mathcal{J}| \geq B$, the affine subspace containing the layer inputs can be estimated by computing the affine span of $\{\bm{u}_j\}_{j \in \mathcal{J}}$.

\subsection{Subspace Estimation}
First, we describe the subspace estimation procedure for the linear subspace induced by the gradients of the FC layers. For each attacked layer $l$, we apply SVD to the observed weight gradient $\bm{G}_{\bm{W}_l}$ and extract an orthonormal basis through the origin, represented by
\[
\bm{P}_l \in \mathbb{R}^{d_{l-1} \times B}.
\]
The SVD is computed with rank $k=B$, corresponding to the batch size.

The affine case is handled similarly. We first construct the ratio-point matrix $\bm{U}$, whose rows correspond to the points $\{\bm{u}_j \mid j \in \mathcal{J}\}$ defined earlier. In practice, to avoid numerical instability caused by near-zero denominators, we define
\[
\mathcal{J} =\left\{ j \in \{1,\ldots,d_l\} \;\middle|\;
|s_j| > 10^{-8} \right\}. 
\]
We then estimate the affine subspace~\footnote{Note that we can estimate affine hulls only for layers with more than $B$ active neurons, i.e., only when $|\mathcal{J}| \geq B$.} using a centered, reliability-weighted SVD with rank $k=B-1$. Each row $u_j$ is assigned the unnormalized weight
\[\tilde{w}_j = |s_j|^2. \]
These weights are normalized over the samples as
\[{w}_j = \frac{\tilde{w_j}}{\sum_{i\in \mathcal{J}} \tilde{w}_i} .\]
The normalized weights determine the contribution of each sample to the SVD fit. Consequently, samples with a larger denominator magnitude, $|s_j|$, exert greater influence on the estimated affine hull. This weighting scheme reduces potential numerical instability introduced by the division and is necessary in the noisy setting.
The resulting affine hull is characterized by
\begin{itemize}
    \item the affine offset $\bm{\mu}_l^\top \in \mathbb{R}^{d_{l-1}}$, 
    \item the orthonormal basis matrix $\bm{P}_l \in \mathbb{R}^{d_{l-1} \times (B-1)}$.
\end{itemize}

\subsection{Subspace Regularizers}
Let $\bm{h}^{(i)}_{l-1}\in\mathbb{R}^{1\times d_{l-1}}$ denote the input activation to layer $l$ for sample $i$. For this layer, let the estimated linear/affine hull be
\[
\mathcal{H}_l=\left\{\bm{\mu}_l + \bm{aP}_l^\top
:\bm{a}\in\mathbb{R}^{1\times k}\right\},
\]
where $\bm{\mu}_l = \bm{0}$, $k=B$ for the linear case and $k=B-1$ for the affine case. Then, the projection of $\bm{h}^{(i)}_{l-1}$ onto $\mathcal{H}_l$ is given by
\[
\Pi_{\mathcal{H}_l}\!\left( \bm{h}^{(i)}_{l-1}\right)=\bm{\mu}_l
+\left(\bm{h}^{(i)}_{l-1}-\bm{\mu}_l\right)\bm{P}_l \bm{P}_l^\top .
\]
Thus, for a batch of activations
\[\bm{H}_{l-1}=\left\{\bm{h}^{(i)}_{l-1}\right\}_{i=1}^B,\]
we define the subspace regularizer at layer $l$ as the average squared distance to that subspace:
\[
\mathcal{R}_{\mathrm{sub}}^{(l)}(\bm{H}_{l-1}) :=
\frac{1}{B} \sum_{i=1}^B \left\| \bm{h}^{(i)}_{l-1} - \Pi_{\mathcal{H}_l}\!\left( \bm{h}^{(i)}_{l-1}\right) \right\|^2 ,  
\]
\[\mathcal{R}_{\mathrm{sub}}^{(l)} := \mathcal{R}_{\mathrm{sub}}^{(l)}(\bm{H}_{l-1}).\]
Note that, if no linear/affine hull was estimated for layer $l$ (insufficient rows), we set:
\[
\mathcal{R}_{\mathrm{sub}}^{(l)}(\bm{H}_{l-1}) := 0, \qquad \mathcal{R}_{\mathrm{sub}}^{(l)} := \mathcal{R}_{\mathrm{sub}}^{(l)}(\bm{H}_{l-1}) .
\]
These regularizers penalize deviations of the layer inputs from the learned low-dimensional subspace structures. 
The corresponding hyperparameters will balance gradient matching performance and adherence to that low-rank structure.

\subsection{Objective Function}
We identify two main approaches for adapting the objective function proposed by Geiping et al.~\cite{ig} to incorporate subspace constraints:
\begin{itemize}
    \item Hard Subspace Input Constraint,
    \item Soft Subspace Input Constraint
\end{itemize}

\subsubsection{Hard Subspace Input Constraint}
In the hard subspace constraint setting, we use the estimated linear/affine hull obtained from the first-layer gradients to reparameterize the batch inputs as
\[\bm{x} = \bm{\mu}_1 + \bm{A P}_1^\top,\]
where $\bm{A} \in \mathbb{R}^{B \times k}$ contains the coefficients optimized for each sample and $k$ is the rank used in the SVD. This reparameterization reduces the number of optimization variables from $B \cdot d_0$ to $B\cdot k$. 
The resulting optimization problem, which combines the hard subspace reparameterization with subspace regularizers for the intermediate layers, is:
\begin{equation}\label{eq: hard_fc}
\min_{\bm{A} ,\bm{y}} \; D_{\mathrm{GM}}(\bm{x},\bm{y}) + \lambda_{\mathrm{TV}} \,  \mathcal{R}_{\mathrm{TV}}(\bm{x}) + \lambda_{\mathrm{sub\text{-}inter}} \sum_{l=2}^L
\mathcal{R}_{\mathrm{sub}}^{(l)},
\end{equation}
where $D_{\mathrm{GM}}$ uses the cosine distance metric and $\lambda_{\mathrm{TV}},$ $\lambda_{\mathrm{sub\text{-}inter}} \ge 0$ control the strength of the regularizers.

\subsubsection{Soft Subspace Input Constraint}
In this setting, we use the estimated linear/affine hull from the first-layer gradients as a regularizer rather than reparameterizing the batch inputs. Additionally, we retain the subspace regularizers for the intermediate layers. The resulting optimization problem is
\begin{equation}\label{eq: soft_fc}
\min_{\bm{x}\in [0,1]^{d_0},\, \bm{y}}\; D_{\mathrm{GM}}(\bm{x},\bm{y}) + \lambda_{\mathrm{TV}}\,\mathcal{R}_{\mathrm{TV}}(\bm{x}) + \lambda_{\mathrm{sub\text{-}inp}}\, \mathcal{R}_{\mathrm{sub}}^{(1)}(\bm{x}) + \lambda_{\mathrm{sub\text{-}inter}} \sum_{l=2}^{L}\mathcal{R}_{\mathrm{sub}}^{(l)},
\end{equation}
where $D_{\mathrm{GM}}$ uses the cosine distance metric and $\lambda_{\mathrm{TV}},$ $\lambda_{\mathrm{sub\text{-}inp}} ,\; \lambda_{\mathrm{sub\text{-}inter}}  \ge 0$ control the strength of the regularizers. 

\subsection{Initialization and Projection}
The objectives in Equations~\eqref{eq: hard_fc} and~\eqref{eq: soft_fc} are highly non-convex, and, thus, the objective value to which they converge depends on the initialization of the optimization variables. Therefore, it is meaningful to ensure randomness in the initialization procedure and potentially perform multiple restarts.

In the soft subspace case, the inputs $\bm{x}$ and the label distribution $\bm{y}$ can be initialized based on the standard normal distribution. During optimization, we enforce valid pixel values by projecting $\bm{x}$ onto the box constraint $[0,1]^{d_0}$ at each iteration.

In the hard-subspace setting, however, the optimization variables for the inputs are the coefficients $\bm{A}$. Directly initializing $\bm{A}$ requires selecting an appropriate standard deviation, and it is unclear how to achieve sufficient diversity while ensuring that the corresponding inputs remain within, or close to, the box constraints. Alternatively, we could randomly initialize $\bm{x}$ and project it onto the low-rank subspace to obtain initial coefficients. However, when the batch size is sufficiently small, the large dimensionality gap between the original space and the low-rank subspace causes the projected points to concentrate in a small region.

To obtain initializations that are both diverse and feasible, we use a Hit-and-Run procedure. The algorithm first identifies a feasible starting point and then randomly explores the admissible region. Finally, we select a well-separated subset of these samples using greedy farthest-point sampling.

In the hard-subspace setting, directly projecting $\bm{x}$ onto the box $[0,1]^{d_0}$ may violate the affine-subspace constraint. Thus, we can either enforce the box constraint through a regularization term during optimization or avoid projection entirely and apply it only to the final reconstructed inputs.

%% file: section_files/4.method_cnn.tex
\section{Extension to CNNs} \label{section 4}
Subspace constraints can only be applied to FC layers, and if a reparameterization is desirable, then our previous approach cannot be applied directly to CNNs. However, each version of our proposed algorithm can be extended to CNNs based on feature inversion (FI) as seen in CPA \cite{cpa}, while the soft subspace case can also bypass FI. In particular, we can identify three approaches for extending our algorithm to CNNs, of which the first two utilize FI:
\begin{itemize}
    \item Hard Subspace Input Constraint 
    \item Soft Subspace Input Constraint
    \item Soft Subspace Intermediate Constraint
\end{itemize}

\subsection{Hard Subspace Input Constraint}
Following CPA, we first recover the embedding vector $\bm{z} \in \mathbb{R}^{B\times d}$, which is produced as input to the CNN's first FC layer, and then use this vector in feature inversion (FI). 
To recover $\bm{z}$, we isolate the fully connected component of the CNN and treat it as a separate model on which the attack is performed. Because the target variable is now an embedding rather than an image, the total variation (TV) prior is no longer applicable. Instead, we can use the sparsity and non-negativity properties of embeddings produced by ReLU activations. These properties are encouraged using the following regularizers: 
\[
\mathcal{R}_{\ell_1}(\bm{z}) = \frac{1}{Bd} \sum_{b=1}^{B} \sum_{j=1}^{d}
\left| z_{b,j} \right|\]
and
\[
\mathcal{R}_{\mathrm{nv}}(\bm{z}) =\frac{1}{B}\sum_{b=1}^{B}
\left\| \operatorname{ReLU}(-\bm{z}_{b,:})\right\|
\]
The $\ell_1$ regularizer promotes sparsity, whereas
$\mathcal{R}_{\mathrm{nv}}$ penalizes negative entries, encouraging the
recovered embeddings to remain consistent with the output of a ReLU
activation.

Additionally, drawing inspiration from decorrelation objectives used in blind source separation (BSS) and CPA, we utilize a milder decorrelation regularizer based on squared cosine similarity. This term encourages distinct recovered embeddings to be approximately orthogonal without imposing an overly aggressive separation constraint. This prior is useful for the embedding reconstruction since the original embeddings are high-dimensional and tend to have low pairwise correlation. The regularizer is defined as
\[
\mathcal{R}_{\mathrm{decor}}(\bm{z})=\mathbb{E}_{i \neq j}
\left[\operatorname{CS}(\bm{z}_i,\bm{z}_j)^2\right],
\]
where
\[
\operatorname{CS}\left(\bm{z}_{i},\bm{z}_{j}\right) = \frac{\bm{z}_{i}^{\top}\bm{z}_{j}}{\left\|\bm{z}_{i}\right\| \left\|\bm{z}_{j}\right\|}.
\]
Thus, the embedding-recovery problem is formulated as
\begin{equation}\label{eq: hard_cnn}
\begin{aligned}
\min_{\bm{A} , \bm{y}} \quad
& D_{\mathrm{GM}}(\bm{z},\bm{y})+ \lambda_{\mathrm{nv}} \, \mathcal{R}_{\mathrm{nv}}(\bm{z}) + \lambda_{\mathrm{\ell_1}} \, \mathcal{R}_{\ell_1}(\bm{z}) \\
& + \lambda_{\mathrm{decor}} \, \mathcal{R}_{\mathrm{decor}}(\bm{z}) +\lambda_{\mathrm{sub\text{-}inter}} \, \sum_{l =2}^L
\mathcal{R}_{\mathrm{sub}}^{(l)},
\end{aligned}
\end{equation}
where $\bm{A}$ are the optimization variables in coefficient space, $D_{\mathrm{GM}}$ uses the cosine distance metric for the corresponding gradients of the FC component only, and $\lambda_{\mathrm{nv}},\;$ $\lambda_{{\ell_1}} ,\;$ $\lambda_{\mathrm{decor}},\;$ $ \lambda_{\mathrm{sub\text{-}inter}}  \ge 0$ control the strength of the regularizers.

\subsection{Soft Subspace Input Constraint}
In this setting, we follow a similar approach by recovering the embedding vector $\bm{z}$. However, we optimize directly in the embedding space and use the estimated linear/affine hull from the first FC-layer gradients as a regularizer. The resulting optimization problem is
\begin{equation}\label{eq: soft_inp_cnn}
\begin{aligned}
\min_{\bm{z}\in \mathbb{R}_+ , \;\bm{y}} \;
& D_{\mathrm{GM}}(\bm{z}, \bm{y}) + \lambda_{\mathrm{\ell_1}} \, \mathcal{R}_{\ell_1}(\bm{z}) + \lambda_{\mathrm{decor}} \, \mathcal{R}_{\mathrm{decor}}(\bm{z})\\
& + \lambda_{\mathrm{sub\text{-}inp}}\, \mathcal{R}_{\mathrm{sub}}^{(1)}(\bm{z}) +\lambda_{\mathrm{sub\text{-}inter}} \, \sum_{l =2}^L
\mathcal{R}_{\mathrm{sub}}^{(l)},
\end{aligned}
\end{equation}
where $D_{\mathrm{GM}}$ uses the cosine distance metric and $\lambda_{{\ell_1}} ,\;$ $\lambda_{\mathrm{decor}},\;$ $\lambda_{\mathrm{sub\text{-}inp}} ,\; \lambda_{\mathrm{sub\text{-}inter}}  \ge 0$ control the strength of the regularizers. The constraint $\bm{z}\in\mathbb{R}_+$ is enforced by projecting the optimized embedding onto the positive orthant at each iteration. This projection-based constraint is well suited to the noiseless setting. In the presence of noise, however, it is preferable to relax the constraint by removing the projection and reintroducing the non-negativity regularizer.

\subsection{Soft Subspace Intermediate Constraint}
Lastly, we consider an alternative approach that bypasses FI and optimizes directly in the input space $\bm{x}$. In particular, the attack can incorporate subspace regularization at the fully connected layers of the model while optimizing over the image pixel space, leading to the optimization problem
\begin{equation}\label{eq: soft_inter_cnn}
\min_{\bm{x} \in [0,1]^{d_0}, \; \bm{y}} D_{\mathrm{GM}} (\bm{x},\bm{y}) + \lambda_{\mathrm{TV}} \, \mathcal{R}_{\mathrm{TV}}(\bm{x}) + 
\lambda_{\mathrm{sub\text{-}inter}} \, \sum_{l \in I}
\mathcal{R}_{\mathrm{sub}}^{(l)},
\end{equation}
where $D_{\mathrm{GM}}$ uses the cosine distance metric over the gradients of the whole model. The hyperparameters
$\lambda_{\mathrm{TV}},\;$ $\lambda_{\mathrm{sub\text{-}inter}} \ge 0$
control the strengths of the regularizers, and
\[
I = \{ i \in \{1,\dots,L\} \mid \text{layer } i \text{ is fully connected} \}.
\]
Although this formulation provides a viable alternative, it has some important limitations. First, the subspace information is obtained from layers located far from the input, making it more difficult for the regularizers to guide the optimization toward the original input. Second, the formulation does not address the inherent difficulties of gradient matching for high-resolution images and deep networks, for which the objective $D_{\mathrm{GM}}$ can be particularly challenging to optimize. In contrast, the two previous approaches first optimize over a smaller portion of the model and utilize the subspace information more effectively, which can lead to better results. For these reasons, we will not focus on this approach.

\subsection{Initialization}
The label distribution $\bm{y}$ is initialized from a standard normal distribution in all settings. To account for the non-negativity of the embeddings, $\bm{z}$ is initialized from a uniform distribution over $[0,1]$. In the hard-subspace setting, the initialized embeddings are subsequently projected onto the low-rank subspace, from which the corresponding initial coefficients $\bm{A}$ are obtained.

In this case, we do not employ the previously introduced Hit-and-Run procedure, since the embeddings are not constrained to a bounded domain. Consequently, the batch members may be initialized at identical or nearby points, limiting the effectiveness of randomized restarts. Nevertheless, the decorrelation regularizer encourages the optimized embeddings to diverge during optimization, reducing the impact of their similar initializations.

%% file: section_files/5.experiments.tex
\section{Experiments}\label{section 5}
\subsection{Experimental Setup}
We evaluate our approaches with both the new affine and the known linear low-rank structure. We first define the experimental setup, including the target models and the baseline attacks against which we compare.

\subsubsection{Models and Datasets} 
We evaluate our attack on both fully connected (FC) and convolutional neural network (CNN) models trained for image classification. For the FC experiments, we use an $L$-layer network, denoted FC-$L$, consisting of $(L-1)$ Linear(512)–Activation blocks followed by a Linear($k$) output layer for $k$-class classification. We focus on the FC-$3$ architecture and consider two activation functions, ReLU and LeakyReLU. In the main results with accurate batch-size information, we report performance for both activation functions. For experiments involving inaccurate batch-size information or the noisy setting, we restrict our evaluation to ReLU.

We train the FC-$3$ model on CIFAR-10 and Tiny-ImageNet for 20 epochs using the Adam optimizer with a learning rate of $10^{-3}$. For the CNN experiments, we use a pretrained VGG-16 model from torchvision and evaluate it on ImageNet.

\subsubsection{Evaluation}
We focus on the SGD setting and evaluate the attacks under different assumptions about batch-size availability. Unless stated otherwise, labels are assumed unknown.

We first consider the case in which the true batch size is known and then study robustness to batch-size inaccuracies. Let $B$ denote the true batch size and $\widehat{B}$ its estimate. An inaccurate batch-size estimate affects the entire attack pipeline, including the rank used for subspace estimation and the number of inputs reconstructed during optimization. For $\widehat{B}<B$, we match the reconstructions to the $\widehat{B}$ most similar ground-truth inputs and average the resulting metrics. For $\widehat{B}>B$, we retain the best $B$ reconstructions, match them to the ground-truth batch, and report the corresponding average metrics.

For the CNN setting, we directly compare the reconstructed embeddings. CPA \cite{cpa} has already demonstrated the effectiveness of combining embedding reconstruction with feature inversion and GMA. Therefore, a direct comparison of the reconstructed latent representation, $\bm{z}$, provides a more meaningful evaluation. We evaluate each attack over 5 data batches and report the average LPIPS score for image reconstruction using the FC-$3$ model, and the average cosine similarity (CS) between the reconstructed and ground-truth embeddings for the VGG-16 model.

All attacks are run for 25K optimization iterations and, when applicable, with two independent initializations. We report the iterate achieving the lowest objective value across all iterations and initializations. CPA is excluded from the multiple-initialization setting since its initialization is deterministic. 

\subsubsection{Baselines}
As discussed earlier, we focus on the honest-but-curious setting, in which in-distribution data are not available, and batch normalization (BN) is not used. We therefore exclude methods that rely on generative models, BN statistics, or an active adversary. The most relevant baselines for our setting are IG~\cite{ig} and CPA~\cite{cpa} for image data.

Although SPEAR and SPEAR++~\cite{spear,spear++} are applicable in some of these settings, we primarily consider approximate, optimization-based attacks that can be extended to arbitrary activation functions. This choice is particularly meaningful for evaluating the threat posed in settings where exact gradient-inversion methods are not applicable. For each baseline, we use the best-performing hyperparameter configuration reported in the corresponding original work. Further hyperparameter details are provided in Appendix \ref{hyperparameters}.

\subsubsection{Hardware}
All experiments were conducted in Google Colab. Experiments involving fully connected models were run on an NVIDIA L4 GPU, whereas experiments with the VGG-16 convolutional model were performed on an NVIDIA A100 GPU.

\subsection{Main Results}

\subsubsection{Accurate Batch Size}

% CIFAR-10 figure for the FC-3 model with batch size 16
\begin{figure}[tb]
	\centering
    \includegraphics[width=\textwidth]{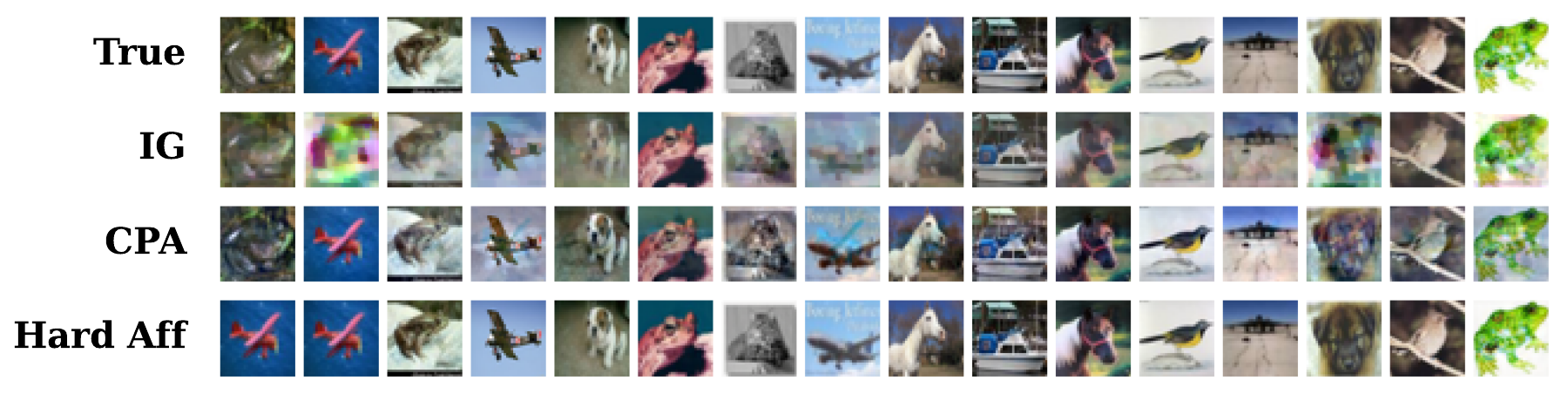}
	\caption{Comparison of a batch of 16 CIFAR-10 images recovered from IG, CPA and Hard Affine attacks for the FC-3 model with ReLU. The quality of images recovered by the Hard Affine variant is better than the baselines but produces a duplicate image.}
	\label{fig:cifar-b16}
\end{figure}

% CIFAR-10 table for the FC-3 model
\begin{table}[tb]
\centering
\caption{LPIPS $\downarrow$ scores on CIFAR-10 for the FC-3 model with accurate batch size.}
\label{tab:fc3-cifar10}

\setlength{\tabcolsep}{4pt}

\begin{tabular}{
  @{}
  l
  @{\hspace{10pt}}
  *{5}{S[table-format=1.3]}
  @{\hspace{20pt}}
  *{5}{S[table-format=1.3]}
  @{}
}
\toprule
& \multicolumn{5}{c}{\textbf{ReLU}}
& \multicolumn{5}{c}{\textbf{LeakyReLU}} \\
\cmidrule(lr{20pt}){2-6}
\cmidrule(lr){7-11}

\textbf{Attack}
& {\textbf{16}}
& {\textbf{32}}
& {\textbf{64}}
& {\textbf{128}}
& {\textbf{256}}
& {\textbf{16}}
& {\textbf{32}}
& {\textbf{64}}
& {\textbf{128}}
& {\textbf{256}} \\
\midrule

IG
& 0.206 & 0.350 & 0.482 & 0.565 & 0.597
& 0.153 & 0.252 & 0.481 & 0.574 & 0.599 \\

CPA
& 0.162 & 0.205 & 0.286 & 0.517 & {\bfseries 0.577}
& 0.163 & 0.210 & 0.305 & 0.523 & {\bfseries 0.580} \\

Soft Linear
& {\bfseries 0.077} & 0.135 & 0.358 & 0.504 & 0.580
& 0.107 & 0.115 & 0.263 & 0.516 & 0.585 \\

Hard Linear
& 0.135 & 0.277 & 0.450 & 0.544 & 0.587
& {\bfseries 0.057} & 0.216 & 0.448 & 0.577 & 0.597 \\

Soft Affine
& 0.161 & 0.137 & {\bfseries 0.206} & {\bfseries 0.474} & 0.587
& 0.087 & {\bfseries 0.087} & {\bfseries 0.139}
& {\bfseries 0.481} & 0.581 \\

Hard Affine
& 0.119 & {\bfseries 0.118} & 0.307 & 0.484 & 0.587
& 0.089 & 0.088 & 0.216 & 0.520 & 0.595 \\

\bottomrule
\end{tabular}
\end{table}

\textbf{Results for FC-3.} 
We first present results for FC-$3$ models with ReLU and LeakyReLU activations trained on CIFAR-10 and Tiny-ImageNet. Tables \ref{tab:fc3-cifar10} and \ref{tab:fc3-tiny-imagenet} report the LPIPS scores comparing the baseline methods with the proposed subspace approaches when the true batch size is known, for batch sizes in $\{16, 32, 64, 128, 256\}$. 

As shown in Table \ref{tab:fc3-cifar10}, the proposed subspace methods exhibit similar behavior for both ReLU and LeakyReLU activations, with the choice of activation function not meaningfully affecting the relative performance of the evaluated methods. Across both settings, the subspace methods can outperform the baselines on CIFAR-10 for batch sizes up to $128$. At a batch size of $16$, the Soft Linear variant performs best for the ReLU case, whereas for batch sizes of $64$ and $128$, the Soft Affine variant maintains an advantage over the baselines under both activation functions, unlike the linear variants in some cases. At smaller batch sizes, the subspace methods also tend to reconstruct images with higher visual fidelity than the baselines, although they may become trapped in local minima and produce duplicate reconstructions. In contrast, CPA tends to recover a larger number of distinct images more consistently, but with lower reconstruction quality, which can lead to higher (i.e., worse) LPIPS scores. An example of this behavior is shown in Figure \ref{fig:cifar-b16}. As discussed in the next section, this limitation can be mitigated by overestimating the batch size.

% Tiny-ImageNet figure for the FC-3 model with LeakyReLU and batch size 128
\begin{figure}[tb]
	\centering
    \includegraphics[width=\textwidth]{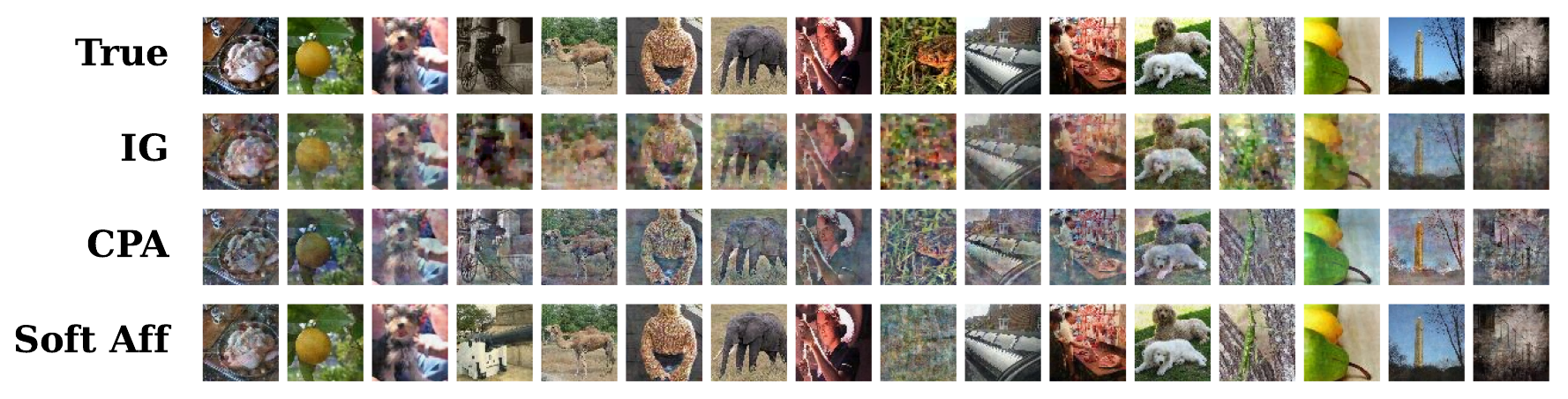}
	\caption{Comparison of a subset of a batch of 128 Tiny-ImageNet images recovered from IG, CPA and Soft Affine attacks for the FC-3 model with LeakyReLU. The quality of most images recovered by the Soft Affine variant is significantly better than the baselines.}
	\label{fig:tiny-imagenet-b128}
\end{figure}

% Tiny-ImageNet table for the FC-3 model
\begin{table}[tb]
\centering
\caption{LPIPS $\downarrow$ scores on Tiny-ImageNet for the FC-3 model with accurate batch size.}
\label{tab:fc3-tiny-imagenet}

\setlength{\tabcolsep}{4pt}
\renewcommand{\arraystretch}{0.9}

\begin{tabular}{
  @{}
  l
  @{\hspace{10pt}}
  *{5}{S[table-format=1.3]}
  @{\hspace{20pt}}
  *{5}{S[table-format=1.3]}
  @{}
}
\toprule
& \multicolumn{5}{c}{\textbf{ReLU}}
& \multicolumn{5}{c}{\textbf{LeakyReLU}} \\
\cmidrule(lr{20pt}){2-6}
\cmidrule(lr){7-11}

\textbf{Attack} 
& {\textbf{16}}
& {\textbf{32}}
& {\textbf{64}}
& {\textbf{128}}
& {\textbf{256}}
& {\textbf{16}}
& {\textbf{32}}
& {\textbf{64}}
& {\textbf{128}}
& {\textbf{256}} \\
\midrule

IG
& 0.469 & 0.634 & 0.703 & 0.726 & 0.739
& 0.115 & 0.192 & 0.329 & 0.406 & 0.556 \\

CPA
& 0.156 & {\bfseries 0.182} & {\bfseries 0.230}
& {\bfseries 0.275} & {\bfseries 0.387}
& 0.156 & 0.182 & 0.229 & 0.274 & 0.394 \\

Soft Linear
& {\bfseries 0.083} & 0.616 & 0.682 & 0.701 & 0.704
& {\bfseries 0.030} & {\bfseries 0.037} & 0.068
& 0.127 & 0.355 \\

Hard Linear
& 0.116 & 0.688 & 0.768 & 0.770 & 0.728
& 0.037 & 0.053 & 0.135 & 0.263 & 0.537 \\

Soft Affine
& 0.186 & 0.366 & 0.563 & 0.664 & 0.685
& 0.054 & 0.062 & {\bfseries 0.067} & {\bfseries 0.092}
& {\bfseries 0.331} \\

Hard Affine
& 0.149 & 0.474 & 0.597 & 0.681 & 0.690
& 0.067 & 0.102 & 0.087 & 0.174 & 0.599 \\

\bottomrule
\end{tabular}
\end{table}

Table \ref{tab:fc3-tiny-imagenet} presents the results on Tiny-ImageNet, where the choice of activation function has a more meaningful effect on performance. With ReLU activations, the subspace-based methods perform substantially worse than CPA for batch sizes of $32$ and above. This behavior can be attributed to the very small number of active neurons in the second layer ($\leq 10$), which falls below the batch size. As a result, the corresponding linear/affine hull cannot be estimated reliably, limiting the amount of useful information available to the subspace-based attacks. In contrast, LeakyReLU avoids the issue of inactive neurons, allowing linear/affine hulls to be estimated from all layers. In this setting, the Soft Linear and Soft Affine variants are particularly effective and consistently outperform the baselines across all evaluated batch sizes.

\textbf{Results for VGG-16.} 
We then evaluate embedding reconstruction for the VGG-16 model pretrained on ImageNet. Table \ref{tab:vgg-16 known-unknown} compares CPA with the proposed subspace-based attacks under both unknown and known label settings, assuming that the true batch size is available to the attacker. We consider batch sizes $B\in\{16,32,64,128,256\}$. Although access to the ground-truth labels may be unrealistic, an attacker may be able to infer them in some cases. More importantly, the known-label setting provides a useful controlled comparison for assessing how strongly label information can improve the proposed attacks in this setting. In contrast, CPA does not use label information, and its performance is therefore identical in the known and unknown label cases.

When labels are unknown, CPA achieves the highest reconstruction score for all batch sizes except $B=16$, where Hard Affine performs best. The performance of all methods generally decreases as the batch size grows, but this degradation is substantially more pronounced for the proposed methods in the unknown-label setting. Providing the true labels substantially improves the subspace-based attacks, especially the hard variants. Hard Linear outperforms CPA across all batch sizes, achieving scores above $0.99$ up to $B=64$ and remaining strong at $B=128$ and $256$ with $0.982$ and $0.957$, respectively. Soft Affine and Hard Affine perform best at $B=16$, reaching $1.000$, but degrade more rapidly as the batch size increases. In contrast, Soft Linear benefits only modestly from label information and remains below CPA across all batch sizes. Overall, label information is particularly beneficial to most subspace methods, with Hard Linear showing the greatest robustness to increasing batch size.

\begin{table}[tb]
\centering
\caption{
CS $\uparrow$ scores on embeddings of ImageNet for VGG-16 with and without known labels. Columns are grouped by the batch size $B$; within each group, we compare attacks without and with access to the ground-truth labels.
}
\label{tab:vgg-16 known-unknown}

\setlength{\tabcolsep}{2.85pt}
\renewcommand{\arraystretch}{1.1}

\begin{tabular}{
    @{}
    l
    @{\hspace{6pt}}
    *{10}{S[table-format=1.3]}
    @{}
}
\toprule
& \multicolumn{2}{c}{$\bm{B=16}$}
& \multicolumn{2}{c}{$\bm{B=32}$}
& \multicolumn{2}{c}{$\bm{B=64}$}
& \multicolumn{2}{c}{$\bm{B=128}$}
& \multicolumn{2}{c}{$\bm{B=256}$} \\
\cmidrule(lr){2-3}
\cmidrule(lr){4-5}
\cmidrule(lr){6-7}
\cmidrule(lr){8-9}
\cmidrule(lr){10-11}

\textbf{Attack}
& {\footnotesize Unknown}
& {\footnotesize Known }
& {\footnotesize Unknown}
& {\footnotesize Known}
& {\footnotesize Unknown}
& {\footnotesize Known}
& {\footnotesize Unknown}
& {\footnotesize Known} 
& {\footnotesize Unknown}
& {\footnotesize Known} \\
\midrule

CPA
& 0.948 & 0.948
& {\bfseries 0.921} & 0.921
& {\bfseries 0.922} & 0.922
& {\bfseries 0.933} & 0.933
& {\bfseries 0.919} & 0.919 \\

Soft Linear
& 0.776 & 0.751
& 0.746 & 0.717
& 0.741 & 0.738
& 0.691 & 0.728 
& 0.658 & 0.724  \\

Hard Linear
& 0.922 & 0.998
& 0.863 & {\bfseries 0.992}
& 0.791 & {\bfseries 0.991}
& 0.676 &{\bfseries 0.982}
& 0.614 & {\bfseries 0.957} \\

Soft Affine
& 0.879 & {\bfseries 1.000}
& 0.777 & 0.864
& 0.727 & 0.908
& 0.621 & 0.732
& 0.576 & 0.702 \\

Hard Affine
& {\bfseries 0.964} & {\bfseries 1.000}
& 0.888 & 0.923
& 0.792 & 0.926
& 0.678 & 0.761
& 0.622 & 0.723 \\

\bottomrule
\end{tabular}
\end{table}

\subsubsection{Inaccurate Batch Size}

\begin{table}[tb]
\centering
\caption{
LPIPS $\downarrow$ scores on CIFAR-10 for the FC-3 model under inexact batch-size assumptions. Columns are grouped by the true batch size $B$; within each group,
the attack assumes batch size $\widehat{B}$.
}
\label{tab:fc3-inexact-batch-size}

\setlength{\tabcolsep}{5pt}
\renewcommand{\arraystretch}{1.1}

\begin{tabular}{
    @{}
    l
    @{\hspace{8pt}}
    *{8}{S[table-format=1.3]}
    @{}
}
\toprule
& \multicolumn{2}{c}{\textbf{True} $\bm{B=16}$}
& \multicolumn{2}{c}{\textbf{True} $\bm{B=32}$}
& \multicolumn{2}{c}{\textbf{True} $\bm{{B=64}}$}
& \multicolumn{2}{c}{\textbf{True} $\bm{B=128}$} \\
\cmidrule(lr){2-3}
\cmidrule(lr){4-5}
\cmidrule(lr){6-7}
\cmidrule(lr){8-9}

\textbf{Attack}
& {$\widehat{B}=12$}
& {$\widehat{B}=32$}
& {$\widehat{B}=26$}
& {$\widehat{B}=64$}
& {$\widehat{B}=56$}
& {$\widehat{B}=96$}
& {$\widehat{B}=112$}
& {$\widehat{B}=160$} \\
\midrule

IG
& 0.292
& 0.155
& 0.374
& 0.288
& 0.491
& 0.452
& 0.557
& 0.555 \\

CPA
& 0.192
& 0.162
& 0.231
& 0.202
& 0.299
& 0.276
& 0.525
& 0.501 \\

Soft Linear
& 0.307
& 0.053
& 0.327
& 0.059
& 0.377
& 0.281
& 0.506
& 0.496 \\

Hard Linear
& 0.365
& 0.088
& 0.400
& 0.109
& 0.467
& 0.373
& 0.535
& 0.521 \\

Soft Affine
& 0.236
& {\bfseries 0.034}
& 0.221
& {\bfseries 0.043}
& {\bfseries 0.261}
& {\bfseries 0.200}
& {\bfseries 0.452}
& {\bfseries 0.483} \\

Hard Affine
& {\bfseries 0.161}
& 0.042
& {\bfseries 0.211}
& 0.045
& 0.318
& 0.209
& 0.470
& 0.505 \\

\bottomrule
\end{tabular}
\end{table}

% vgg-16
\begin{table}[tb]
\centering
\caption{
CS $\uparrow$ scores on embeddings of ImageNet for VGG-16 without known labels under the overestimated batch-size assumption. For each true batch size $B$,
the attack assumes a larger batch size $\widehat{B}$.
}
\label{tab:overestimated-batch-size-vgg16}

\setlength{\tabcolsep}{8pt}
\renewcommand{\arraystretch}{1.1}

\begin{tabular}{
    @{}
    l
    @{\hspace{8pt}}
    *{5}{S[table-format=1.3]}
    @{}
}
\toprule
\textbf{Attack}
& {\shortstack{$\bm{B=16}$\\$\widehat{B}=32$}}
& {\shortstack{$\bm{B=32}$\\$\widehat{B}=64$}}
& {\shortstack{$\bm{B=64}$\\$\widehat{B}=96$}}
& {\shortstack{$\bm{B=128}$\\$\widehat{B}=160$}}
& {\shortstack{$\bm{B=256}$\\$\widehat{B}=300$}} \\
\midrule

CPA
& 0.975
& 0.963
& {\bfseries 0.950}
& {\bfseries 0.952}
& {\bfseries 0.918} \\

Soft Linear
& 0.883
& 0.874
& 0.788
& 0.727
& 0.695 \\

Hard Linear
& {\bfseries 0.999}
& {\bfseries 0.985}
& 0.880
& 0.717
& 0.631 \\

Soft Affine
& 0.979
& 0.954
& 0.829
& 0.740
& 0.702 \\

Hard Affine
& 0.991
& 0.965
& 0.846
& 0.705
& 0.634 \\

\bottomrule
\end{tabular}
\end{table}

To assess the sensitivity of the reconstruction algorithms to batch-size inaccuracies, we evaluate image reconstruction for the FC-3 ReLU model trained on CIFAR-10 and embedding reconstruction for VGG-16 pretrained on ImageNet.

\textbf{Results for FC-3.} 
Table \ref{tab:fc3-inexact-batch-size} reports the FC-3 results when the attacker either underestimates or overestimates the true batch size. Comparing these results with the exact-batch setting in Table \ref{tab:fc3-cifar10} shows that underestimating the batch size generally degrades reconstruction quality across all methods. For $B\in\{16,32,64\}$, even a relatively small underestimate consistently increases LPIPS. At $B=128$, several methods appear to improve when $\hat{B}=112$; however, this comparison is biased because only $112$ reconstructions are evaluated. When compared instead with the best $112$ reconstructions obtained under the exact-batch setting, underestimating the batch size again leads to worse performance.

Overestimation is less harmful and often improves reconstruction quality, particularly for the subspace-based methods at small and moderate batch sizes. For example, at $B=16$, Soft Affine improves from $0.161$ with the exact batch size to $0.034$ when $\hat{B}=32$, while Soft Linear improves from $0.077$ to $0.053$. Similar gains are observed at $B=32$ and $64$. These improvements are largely because the additional reconstruction slots reduce the impact of duplicate solutions and increase the chance of recovering all samples in the batch. The extra outputs, however, are mostly duplicates for the subspace methods and noise for CPA due to its decorrelation constraint. CPA shows little benefit from overestimation at $B=16$ and $32$, with larger gains appearing at $B=64$ and $128$. Overall, the results show a clear asymmetry: underestimating the batch size is consistently detrimental, whereas moderate overestimation is generally more robust and can even improve reconstruction quality.

\textbf{Results for VGG-16.} 
For VGG-16, we evaluate only the overestimated batch-size setting, since the FC-3 experiments already show that underestimation consistently degrades reconstruction quality. 

Table \ref{tab:overestimated-batch-size-vgg16} shows that overestimating the batch size again improves reconstruction performance for nearly all methods compared with the unknown-label exact-batch results in Table \ref{tab:vgg-16 known-unknown}. The gains are generally larger for the subspace-based attacks, particularly Soft Affine. At small batch sizes, this makes the subspace methods highly competitive even without label knowledge, with Hard Linear achieving the best scores at $B=16$ and $32$. However, CPA degrades less as the true batch size increases and achieves the highest scores for $B\geq64$. Thus, while overestimating the batch size can substantially benefit the subspace-based attacks, it does not eliminate their stronger degradation at larger batch sizes.

\subsection{Defenses} \label{section 5.3}
\textbf{Gradient perturbation setup.}
Differential privacy (DP) is a popular approach for mitigating privacy leakage in federated learning by providing formal guarantees on the influence of individual training examples on the learned model \cite{dp_sgd}. In DP-SGD, per-sample gradients are clipped to a predefined norm $C$, after which Gaussian noise is added to the aggregated gradient. This introduces a privacy-utility trade-off, as stronger perturbation can reduce the information contained in the gradients but may also degrade model utility.

In our experiments, we implement the gradient clipping and perturbation mechanism using Opacus \cite{opacus}. We fix the clipping threshold to $C=1$ and vary the Gaussian noise multiplier $\sigma$ to study the robustness of the reconstruction attacks to different levels of gradient perturbation. Importantly, the defense is applied only during gradient collection, and the underlying models are not trained with DP-SGD. This keeps the model parameters and corresponding clean gradients fixed across experiments, allowing the effect of clipping and noise on reconstruction performance to be isolated more directly.

\textbf{SNR calibration.}
We characterize the perturbation level using the signal-to-noise ratio (SNR) between the clipped gradient and the added Gaussian noise. The SNR is computed as
\[
\mathrm{SNR}=10\log_{10}\left(\frac{\|\bm{g}_{\mathrm{clip}}\|_2^2}{\mathbb{E}\| \bm{n} \|_2^2}\right),
\]
where $\bm{g}_{\mathrm{clip}}$ denotes the clipped gradient before noise addition and $\bm{n}$ denotes the Gaussian noise, both concatenated over all model parameters. The expected noise power is determined from the variance of the Gaussian perturbation. Since the same noise multiplier can result in different SNRs across samples and architectures, we select a single value of $\sigma$ for each architecture and target SNR such that the average SNR over the 5 evaluated data batches is close to the desired value. We keep this $\sigma$ fixed across all 5 batches, ensuring a consistent defense configuration and identical DP parameters across reconstructions.

All defense experiments are performed with batch size $B=32$.
For the FC-3 model, we evaluate the ReLU network trained on CIFAR-10 at target SNR levels of $\{-5, 0,5,10, 15\} \, \mathrm{dB}$, while for VGG-16 pretrained on ImageNet we consider $\{-20,-15,-10,-5,0\} \, \mathrm{dB}$.

The corresponding noise multipliers and average achieved SNR values are reported in Table \ref{tab:dp-snr-sigma}. Lower SNR values correspond to stronger gradient perturbation and therefore to a more challenging reconstruction setting. The different SNR ranges for FC-3 and VGG-16 are chosen to cover the transition from successful to degraded reconstruction for each model.

\begin{table}[tb]
    \centering
    \caption{Noise multipliers selected for the target SNR levels in the FC-3 and VGG-16 defense experiments. For each target, $\sigma$ is fixed across the 5 evaluated batches. Measured SNR is reported as mean $\pm$ standard deviation across batches.}
    \label{tab:dp-snr-sigma}

    \setlength{\tabcolsep}{6pt}
    \renewcommand{\arraystretch}{1.1}

    \sisetup{
        table-number-alignment = center,
        uncertainty-mode = separate,
        table-align-uncertainty = true
    }

    \begin{tabular}{
        S[table-format=-2.0]
        S[table-format=-2.3(3)]
        S[table-format=1.2]
        @{\hspace{12pt}}
        S[table-format=-2.0]
        S[table-format=-2.3(3)]
        S[table-format=1.2]
    }
        \toprule
        \multicolumn{3}{c}{\textbf{FC-3}} &
        \multicolumn{3}{c}{\textbf{VGG-16}} \\
        \cmidrule(lr){1-3}
        \cmidrule(lr){4-6}

        \multicolumn{1}{c}{Target SNR} &
        \multicolumn{1}{c}{Measured SNR} &
        \multicolumn{1}{c}{$\sigma$ ($\times 10^{-3}$)} &
        \multicolumn{1}{c}{Target SNR} &
        \multicolumn{1}{c}{Measured SNR} &
        \multicolumn{1}{c}{$\sigma$ ($\times 10^{-3}$)} \\
        
        \multicolumn{1}{c}{(dB)} &
        \multicolumn{1}{c}{(dB)} &
        {} &
        \multicolumn{1}{c}{(dB)} &
        \multicolumn{1}{c}{(dB)} &
        {} \\

        \midrule

         -5 & -4.997 +- 0.740 & 8.17 &
        -20 & -19.992 +- 0.258 & 4.55 \\

         0 & -0.007 +- 0.740 & 4.60 & 
        -15 & -14.996 +- 0.258 & 2.56 \\

          5 &  5.016 +- 0.740 & 2.58 &
        -10 & -9.999  +- 0.258 & 1.44 \\

         10 & 10.021 +- 0.740 & 1.45 &
         -5 & -5.001  +- 0.258 & 0.81 \\

         15 & 14.972 +- 0.740 & 0.82 &
          0 & -0.087  +- 0.258 & 0.46 \\

        \bottomrule
    \end{tabular}
\end{table}

\textbf{Reconstruction post-processing.}
Under strong gradient perturbation, we observe that the subspace-based reconstruction methods for the FC-3 experiments can exhibit a compressed intensity range, giving some images a washed-out appearance. This behavior is specific to the subspace variants and is not observed for the baseline methods. We can therefore apply percentile-based contrast stretching as a post-processing step to the reconstructions produced by the subspace methods. Its effect is image-dependent, ranging from minor adjustments when the reconstruction is already sufficiently visible to more noticeable contrast enhancement when the intensity range is strongly compressed.

For each reconstructed FC-3 batch, we compute the 0.1st and 99.9th percentiles, denoted by $P_{0.1}$ and $P_{99.9}$, over all reconstructed pixel values in the batch. The same percentile bounds are therefore used for all images within a batch, ensuring that they undergo the same intensity transformation. The reconstructed images are then rescaled according to
\[
\hat{\bm{x}}^{+}
=
\operatorname{clip}_{[0,1]}
\left(
\frac{\hat{\bm{x}}-P_{0.1}}
{P_{99.9}-P_{0.1}+\epsilon}
\right),
\]
where $\hat{\bm{x}}$ denotes the reconstructed batch and $\epsilon$ is a small constant for numerical stability. Values below $P_{0.1}$ and above $P_{99.9}$ are clipped to 0 and 1, respectively.

We use percentile bounds rather than the absolute minimum and maximum to reduce sensitivity to a small number of extreme pixel values while otherwise preserving nearly the full reconstructed intensity range. Importantly, this operation is performed only after reconstruction and does not modify the observed gradients or the reconstruction optimization itself. Its purpose is to compensate for the contrast compression observed in the FC-3 outputs of the subspace methods rather than to recover additional information from the gradients. As such, it can be viewed as an optional post-processing step available to an attacker when the recovered images exhibit this washed-out appearance.

In the FC-3 results, a $+$ suffix denotes a subspace reconstruction followed by this percentile-based contrast stretching (e.g., \textit{Soft Affine+}). Since the baselines do not exhibit the same washed-out behavior, no corresponding post-processing is applied to them. In particular, IG already produces batches that span the full $[0,1]$ intensity range, so applying the same transformation would leave their outputs largely unchanged. Additional qualitative and quantitative comparisons of FC-3 reconstructions before and after contrast stretching are provided in Appendix~\ref{post-processing impact}. For the VGG-16 experiments, all reported results correspond to the original reconstruction pipeline without this post-processing step. 

\begin{figure}[tb] 
    \centering
    \begin{subfigure}{.49\textwidth}
        \centering
        \includegraphics[width=1\linewidth]{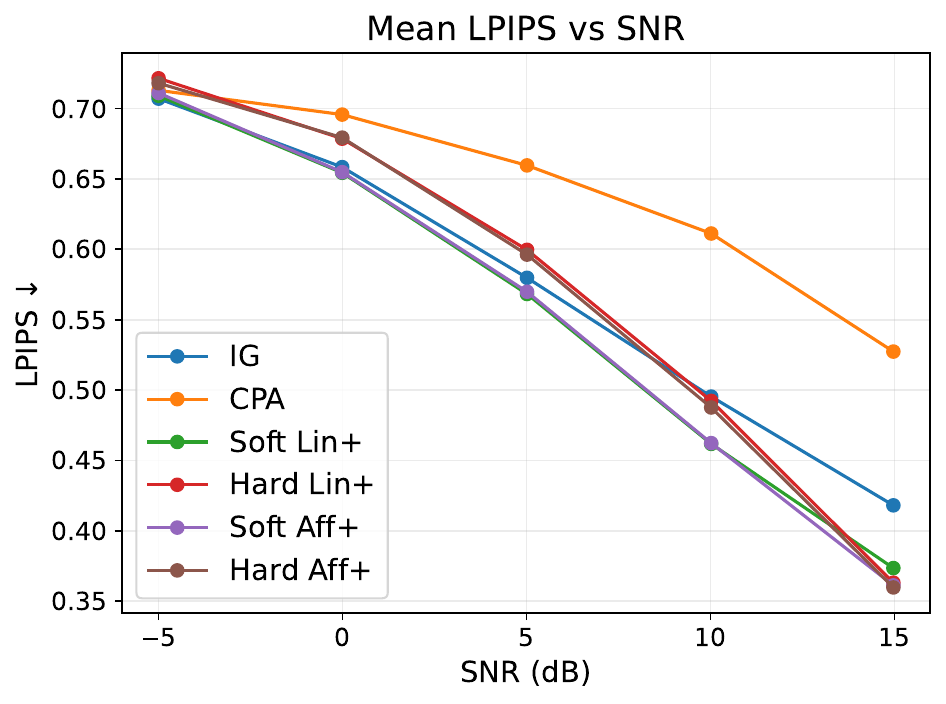}
        \caption{}
        \label{fig:cifar-varying-snr-a}
    \end{subfigure}
    \begin{subfigure}{.49\textwidth} 
        \centering
        \includegraphics[width=1\linewidth]{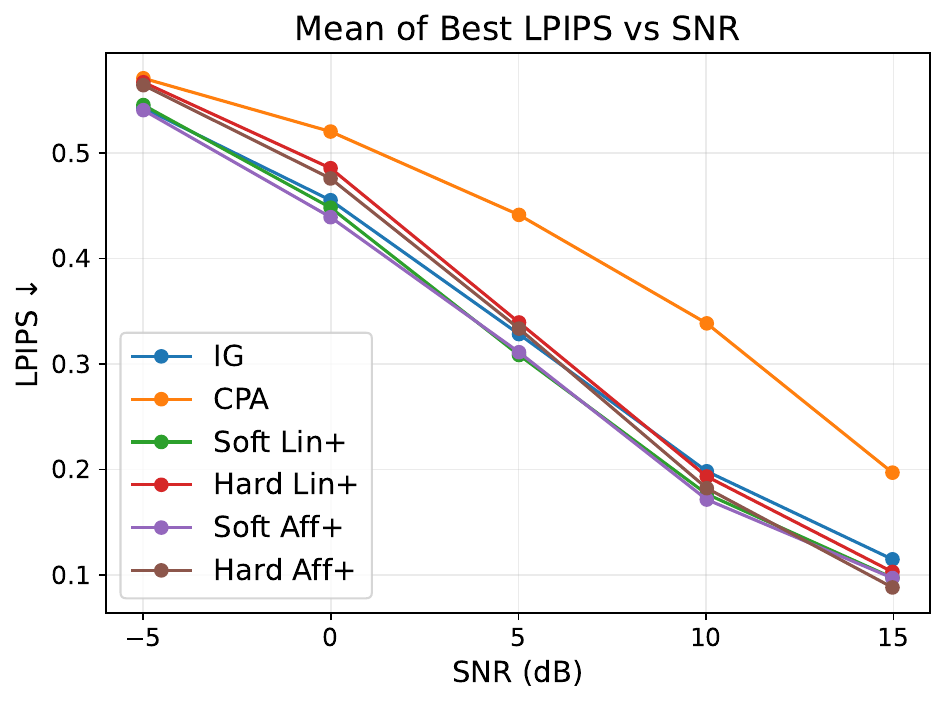}
        \caption{}
        \label{fig:cifar-varying-snr-b}
    \end{subfigure}

    \caption{Reconstruction performance versus SNR for FC-3 with ReLU on CIFAR-10, evaluated over five batches of $B=32$. (a) Mean LPIPS across all reconstructions. (b) Mean LPIPS of the best reconstruction from each batch. Lower values indicate better reconstruction quality.}
    \label{fig:cifar-varying-snr}
\end{figure}
\begin{figure}[tb] 
    \centering
    \begin{subfigure}{.49\textwidth}
        \centering
        \includegraphics[width=1\linewidth]{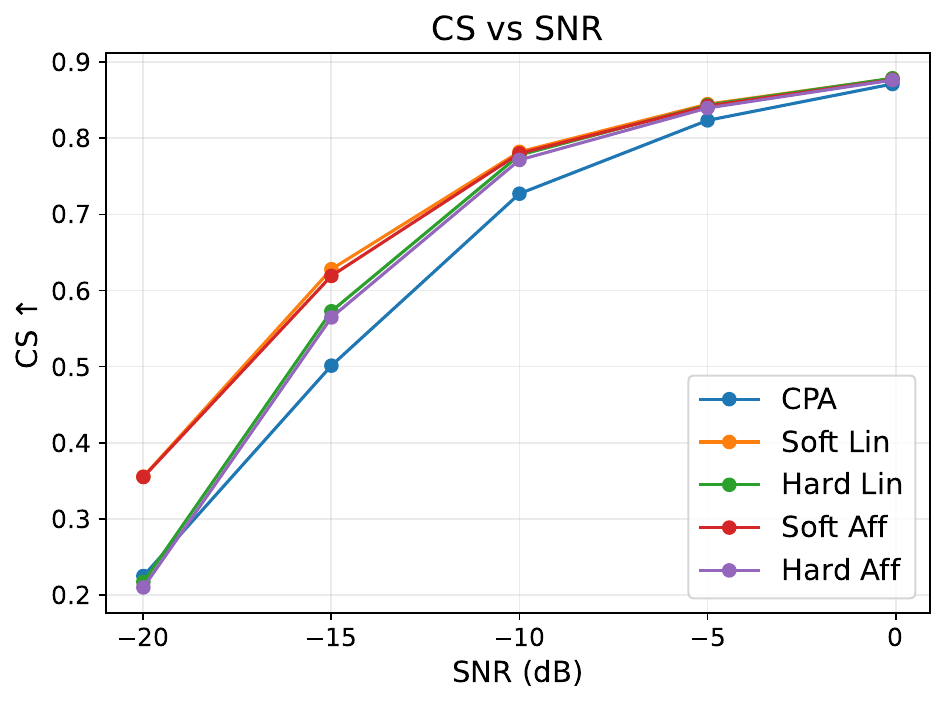}
        \caption{}
        \label{fig:imagenet-varying-snr-a}
    \end{subfigure}
    \begin{subfigure}{.49\textwidth} 
        \centering
        \includegraphics[width=1\linewidth]{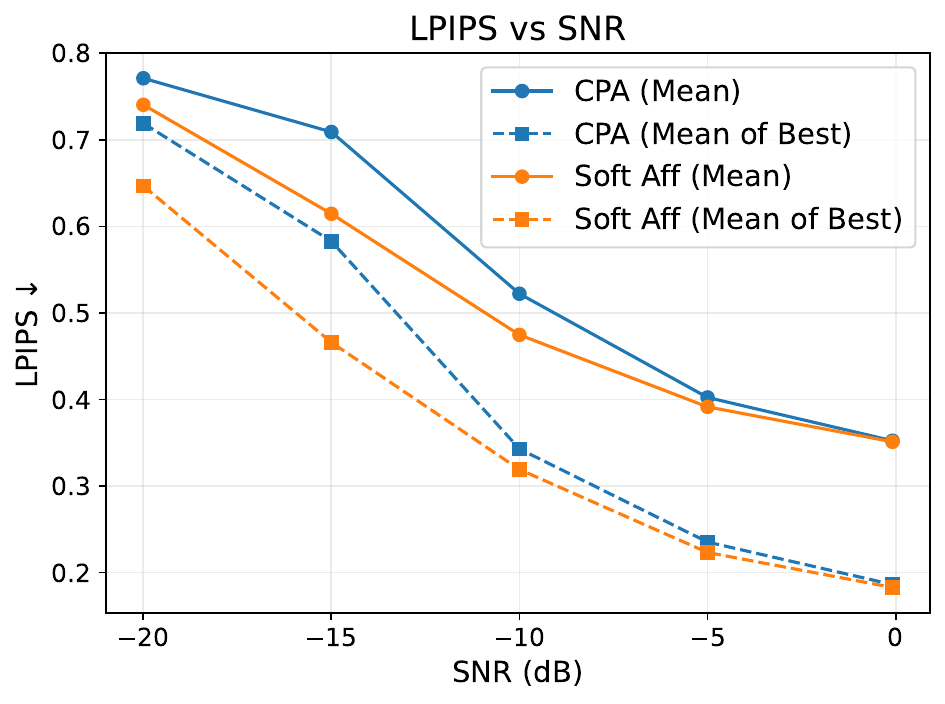}
        \caption{}
        \label{fig:imagenet-varying-snr-b}
    \end{subfigure}

    \caption{Reconstruction performance versus SNR for VGG16 on ImageNet, evaluated over five batches of $B=32$. (a) Mean cosine similarity (CS) between original and reconstructed embeddings. (b) Mean LPIPS and Mean of Best LPIPS computed on the final reconstructed images after feature inversion. Higher CS and lower LPIPS indicate better reconstruction quality.}
    \label{fig:imagenet-varying-snr}
\end{figure}

\textbf{Results under gradient perturbation.}
Figures~\ref{fig:cifar-varying-snr-a} and \ref{fig:imagenet-varying-snr-a} show that the average reconstruction quality improves consistently with SNR for both FC-3 and VGG-16. For FC-3, the differences between methods are relatively small at $-5\,\mathrm{dB}$, where reconstruction is strongly degraded. From $0$ to $15\,\mathrm{dB}$, the soft subspace variants generally achieve lower LPIPS than the baselines, with CPA being the most affected by gradient perturbation and an increasing gap observed relative to IG. For VGG-16, the subspace variants are particularly advantageous at low SNR, where the soft formulations recover embeddings with higher cosine similarity than CPA and the hard variants. The differences among the VGG-16 methods diminish as SNR increases, with all methods approaching similar performance near $0\,\mathrm{dB}$. Interestingly, the choice between soft and hard constraints has a considerably larger effect on reconstruction performance than the choice between the linear and affine formulations, which behave similarly across SNR levels.

To assess whether the VGG-16 embedding trends translate to final image reconstruction, we additionally perform feature inversion on embeddings recovered by CPA and Soft Affine, representing the baseline and a strong subspace variant, respectively. Following CPA, the inversion objective also includes gradient matching, using both the recovered embedding and the observed gradients \cite{cpa}. Figure~\ref{fig:imagenet-varying-snr-b} reports LPIPS across the same SNR levels and shows that Soft Affine consistently outperforms CPA, in line with the embedding reconstruction results. The same trend is observed for the best reconstruction within each batch, with the largest differences occurring at lower SNR and the gap narrowing as the gradients become cleaner.

As indicated by the best-of-batch LPIPS results in Figures~\ref{fig:cifar-varying-snr-b} and \ref{fig:imagenet-varying-snr-b}, and supported by visual inspection, meaningful reconstruction of at least some images remains possible down to approximately $0\,\mathrm{dB}$ for FC-3 and $-15\,\mathrm{dB}$ for VGG-16, respectively. Representative examples are provided in Appendix \ref{additional results}. These SNR levels correspond to noise multipliers of  $4.60\times10^{-3}$ and $2.56\times10^{-3}$. Thus, substantially lower SNR is required to suppress meaningful reconstruction for VGG-16, despite its corresponding noise multiplier being smaller. This highlights the architecture-dependent relationship between gradient magnitude, injected noise, and SNR and shows why the noise multiplier alone is not directly comparable across the two models.

Overall, the results show that the subspace formulations are more robust to gradient perturbation, although the regime in which their advantage is most pronounced differs between the two architectures. For FC-3, the clearest separation from the baselines occurs over $0$ to $15\,\mathrm{dB}$, while at $-5\,\mathrm{dB}$ all methods are strongly degraded. For VGG-16, the advantage is most pronounced at lower SNR and diminishes as the gradients become cleaner.

%% file: section_files/6.limitations_future_work.tex
\section{Limitations and Future Work}
Although the proposed subspace formulations achieve promising reconstruction results, their effectiveness depends on several factors, including the network architecture, batch size, and the availability of label information in certain settings. Further evaluation across different architectures, datasets, and data domains, such as tabular data, would help establish the generality of the findings. Moreover, while affine constraints improve reconstruction over existing linear formulations in certain settings, these improvements are not consistent across all configurations. Investigating the conditions under which affine subspace constraints provide an advantage over linear constraints could help clarify their relative effectiveness.

The evaluation under gradient perturbation is limited to a single batch size $(B=32)$, a fixed clipping threshold, and varying Gaussian noise levels. Although the considered noise multipliers may correspond to weak formal privacy guarantees, future work should investigate the relationship between empirical resistance to reconstruction, formal DP guarantees, and model utility across different privacy configurations. In particular, the strong perturbation required to suppress meaningful reconstruction in some settings raises questions about its impact on model performance and the practical privacy–utility trade-offs of DP mechanisms.

%% file: section_files/7.conclusion.tex
\section{Conclusion}
In this work, we investigated subspace-based formulations for gradient inversion in the image domain, considering linear and affine subspaces under soft and hard constraints. We evaluated their effectiveness across different network architectures, datasets, batch sizes, and attacker assumptions, including settings with gradient perturbations introduced by a DP mechanism.

The experimental results show that our subspace approaches can improve reconstruction quality over existing baselines, although their effectiveness depends on the experimental setting. For FC-3, the best-performing subspace variants outperform CPA on CIFAR-10 at batch sizes up to $128$ under both activations, and on Tiny-ImageNet across all tested batch sizes with LeakyReLU. However, CPA performs best on CIFAR-10 at $B=256$ and on Tiny-ImageNet with ReLU at $B\geq32$. For VGG-16, label availability strongly influences the performance of the subspace methods. Affine constraints offer advantages over their linear counterparts in certain settings, but the two perform similarly under the evaluated DP gradient perturbations at $B=32$, where the subspace approaches generally outperform CPA.

The batch-size experiments reveal that underestimation generally degrades reconstruction quality, whereas moderate overestimation can preserve or even improve it. This asymmetry suggests that precise knowledge of the true batch size is not always necessary, as a moderately overestimated value can suffice. Overall, these findings demonstrate the value of exploiting multi-layer subspace structure for gradient inversion and accounting for uncertainty in batch-size knowledge when assessing privacy leakage from shared gradients.

%% file: section_files/appendix.tex
\appendix
\section{Hyperparameter Settings} \label{hyperparameters}
This appendix reports the hyperparameter settings used throughout the experiments. Separate configurations are provided for the FC and VGG models, as the VGG experiments involve additional hyperparameters.

For the FC models, the same hyperparameter configuration is used in the noiseless setting for both accurate $B$ and overestimated $\widehat{B}$. As reported in Table~\ref{tab:fc_hyperparameters}, the subspace regularization parameters are reduced under the noisy setting, while all other hyperparameters remain unchanged.

For the VGG model, Table~\ref{tab:vgg_hyperparameters} reports the hyperparameter configurations for the noiseless setting with accurate $B$ and overestimated $\widehat{B}$, as well as for the noisy setting. Relative to the accurate-$B$ configuration, the overestimated-$\widehat{B}$ configuration changes only $\lambda_{\mathrm{decor}}$ for Soft Affine, setting it to $0.1$. The noisy setting retains this value and additionally adjusts $\lambda_{\mathrm{nv}}$ for the subspace attacks, as reported in the table. All other hyperparameters remain unchanged.

When the linear and affine variants of a subspace method use identical hyperparameter values, we report them jointly as Soft Subspace or Hard Subspace for conciseness. A dash (--) indicates that the corresponding hyperparameter is not used for that attack. In the noiseless setting, $\lambda_{\mathrm{nv}}$ is not used for the Soft Subspace attacks because non-negativity is enforced through projection onto the positive orthant. In the noisy setting, this projection is removed to relax the non-negativity constraint, which is instead encouraged through the regularization term weighted by $\lambda_{\mathrm{nv}}$.

\begin{table}[tb]
    \centering
    \caption{Hyperparameter settings for the FC models under the noiseless and noisy settings. Only configurations that differ from the preceding setting are reported.}
    \label{tab:fc_hyperparameters}

    \begin{tabular}{llcccccc}
        \toprule
        Setting & Attack
        & $\mathrm{lr}$ 
        & $\lambda_{\mathrm{TV}}$
        & $\lambda_{\mathrm{sub\text{-}inp}}$
        & $\lambda_{\mathrm{sub\text{-}inter}}$
        & $\lambda_{\mathrm{decor}}$
        & $T$ \\
        \midrule

        \multirow{4}{*}{Noiseless}
        & CPA
        & $10^{-3}$ & 3.1 & -- & -- & 1.47 & 12.4 \\

        & IG
        & 0.05 & $10^{-3}$ & -- & -- & -- & -- \\

        & Soft Subspace
        & 0.05 & $10^{-4}$ & $10^{-4}$ & $10^{-4}$ & -- & -- \\

        & Hard Subspace
        & 0.05 & $10^{-4}$ & -- & $10^{-4}$ & -- & -- \\

        \midrule

        \multirow{2}{*}{Noisy}
        & Soft Subspace
        & 0.05 & $10^{-4}$ & $10^{-5}$ & $10^{-5}$ & -- & -- \\

        & Hard Subspace
        & 0.05 & $10^{-4}$ & -- & $10^{-5}$ & -- & -- \\

        \bottomrule
    \end{tabular}
\end{table}

\begin{table}[tb]
    \centering
    \caption{Hyperparameter settings for the VGG model under the noiseless setting with accurate $B$ or overestimated $\widehat{B}$, and under the noisy setting. Only configurations that differ from the preceding setting are reported.}
    \label{tab:vgg_hyperparameters}

    \begin{tabular}{llcccccccc}
        \toprule
        Setting & Attack
        & $\mathrm{lr}$
        & $\lambda_{\mathrm{TV}}$
        & $\lambda_{\mathrm{sub\text{-}inp}}$
        & $\lambda_{\mathrm{sub\text{-}inter}}$
        & $\lambda_{\mathrm{nv}}$
        & $\lambda_{\ell_1}$
        & $\lambda_{\mathrm{decor}}$
        & $T$ \\
        \midrule

        \multirow{4}{*}{Accurate $B$}
        & CPA
        & $10^{-3}$ & 0.1 & -- & -- & 0.13 & 5 & 5.3 & 7.7 \\

        & Soft Affine
        & 0.06 & 0.1 & $5 \cdot 10^{-4}$ & $10^{-4}$
        & -- & $10^{-3}$ & 2 & -- \\

        & Soft Linear
        & 0.06 & 0.1 & $5 \cdot 10^{-4}$ & $10^{-4}$
        & -- & $10^{-3}$ & 0.1 & -- \\

        & Hard Subspace
        & 0.06 & 0.1 & -- & $10^{-4}$
        & 0.1 & $10^{-3}$ & 2 & -- \\

        \midrule

        Overest. $\widehat{B}$
        & Soft Affine
        & 0.06 & 0.1 & $5 \cdot 10^{-4}$ & $10^{-4}$
        & -- & $10^{-3}$ & 0.1 & -- \\

        \midrule

        \multirow{2}{*}{Noisy}
        & Soft Subspace
        & 0.06 & 0.1 & $5 \cdot 10^{-4}$ & $10^{-4}$
        & $10^{-3}$ & $10^{-3}$ & 0.1 & -- \\

        & Hard Subspace
        & 0.06 & 0.1 & -- & $10^{-4}$
        & $10^{-3}$ & $10^{-3}$ & 2 & -- \\

        \bottomrule
    \end{tabular}
\end{table}

\section{Qualitative and Quantitative Effects of Contrast Stretching} \label{post-processing impact}

\begin{figure}[tb] 
    \centering
    \begin{subfigure}{.49\textwidth}
        \centering
        \includegraphics[width=1\linewidth]{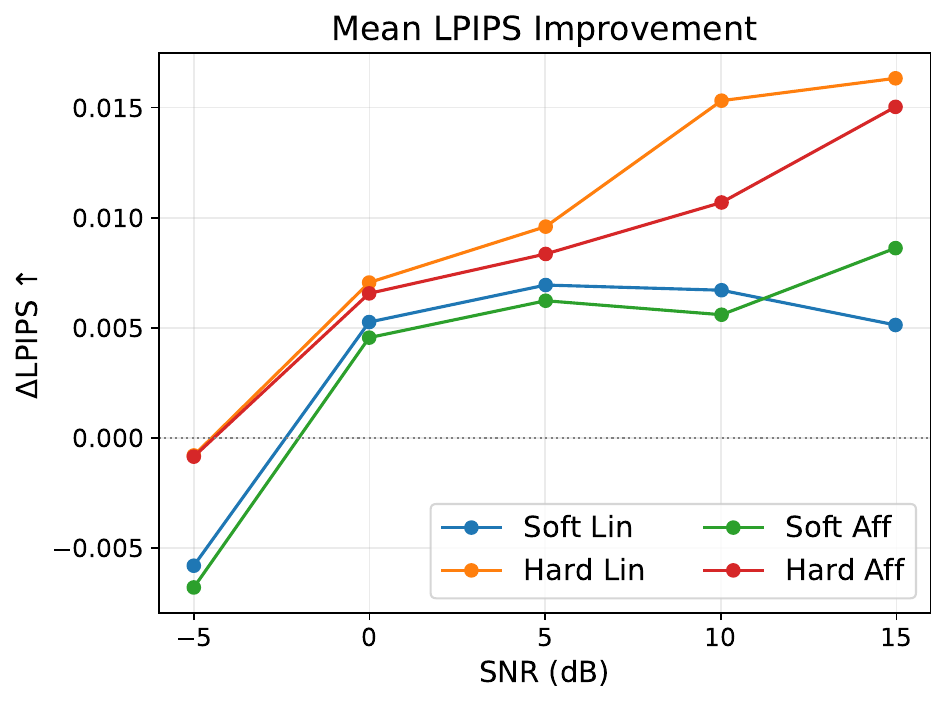}
        \caption{}
        \label{fig:cifar-postprocessing-delta-a}
    \end{subfigure}
    \begin{subfigure}{.49\textwidth} 
        \centering
        \includegraphics[width=1\linewidth]{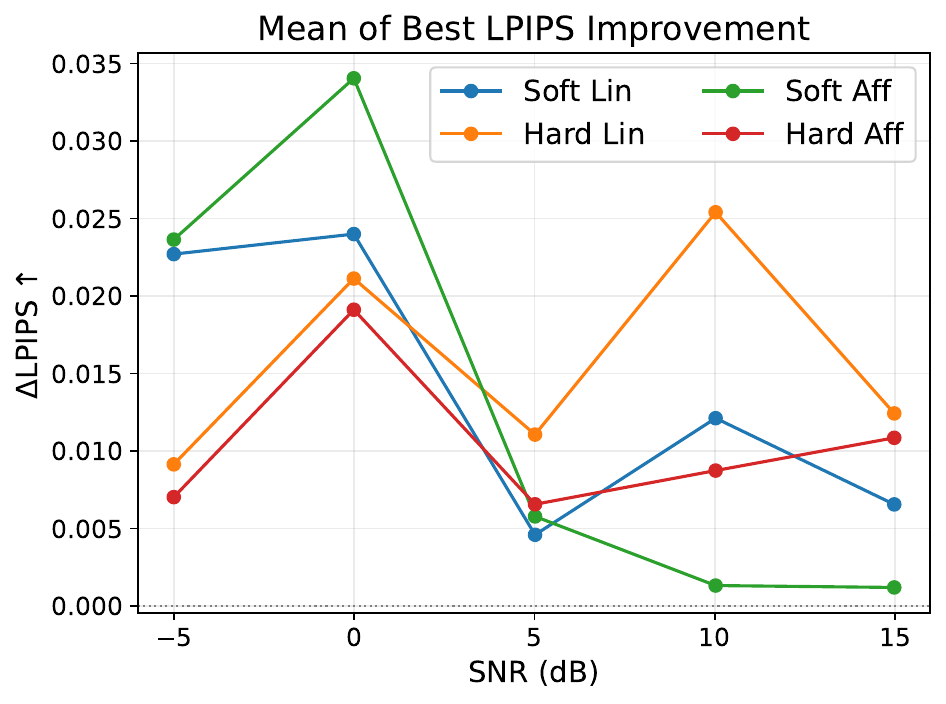}
        \caption{}
        \label{fig:cifar-postprocessing-delta-b}
    \end{subfigure}

    \caption{Effect of percentile-based contrast stretching on the reconstruction quality of the FC-3 model on CIFAR-10 across different SNR levels. (a) Improvement in mean LPIPS. (b) Improvement in mean of best LPIPS. Improvements are computed by subtracting the post-processed scores from the original scores, with positive values indicating better reconstruction quality after contrast stretching.}
    \label{fig:cifar-postprocessing-delta}
\end{figure}

\begin{figure}[tb]
    \centering
    \includegraphics[width=\textwidth]{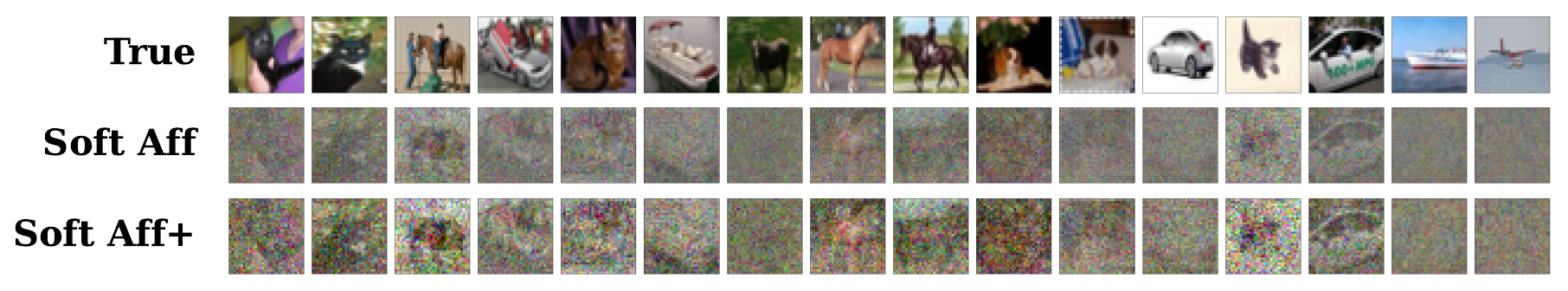}
    \caption{Comparison of CIFAR-10 reconstructions for the FC-3 model before and after percentile-based contrast stretching at approximately $0\,\mathrm{dB}$ SNR. A subset of 16 images from a batch of $B=32$ is shown, comparing the ground-truth images (True), the original Soft Affine reconstructions (Soft Aff), and their post-processed counterparts (Soft Aff+). Percentile bounds are computed over the entire reconstructed batch.}
    \label{fig:cifar-postprocessing-snr-0}
\end{figure}
To assess the effect of the percentile-based contrast stretching described in Section~\ref{section 5.3}, we compare the original FC-3 subspace reconstructions with their post-processed counterparts. Figure~\ref{fig:cifar-postprocessing-delta} reports the change in (a) mean LPIPS across all reconstructions and (b) the mean of the minimum LPIPS values across reconstruction batches, for each evaluated SNR level. The change is computed as the LPIPS score before post-processing minus the score after post-processing, such that positive values indicate improved reconstruction quality.

For mean LPIPS, contrast stretching generally provides small improvements at SNR levels of $0\,\mathrm{dB}$ and above, particularly for the hard-constrained variants, while slightly degrading performance at $-5\,\mathrm{dB}$. In contrast, the mean of best LPIPS improves across all evaluated methods and SNR levels, although the magnitude of the improvement varies. Overall, these results indicate that contrast stretching provides modest quantitative benefits, depending on the reconstruction method, perturbation level, and characteristics of the reconstructed batch.

Figure~\ref{fig:cifar-postprocessing-snr-0} provides a qualitative comparison at approximately $0\,\mathrm{dB}$ SNR, showing 16 images from a batch of $B=32$ reconstructed using Soft Affine, before and after post-processing. For this batch, the percentile bounds are $P_{0.1}=0.211$ and $P_{99.9}=0.756$, indicating that nearly all reconstructed pixel values lie within a substantially compressed intensity range. Following contrast stretching, the mean LPIPS decreases slightly from $0.657$ to $0.646$, while the best LPIPS decreases more substantially from $0.538$ to $0.427$. This batch exhibits the largest improvement in best LPIPS among the evaluated Soft Affine batches at this SNR level and therefore illustrates a particularly favorable case for post-processing. Visually, contrast stretching makes certain image structures and colors more distinguishable, although it can also amplify existing noise and reconstruction artifacts. Thus, the post-processing can improve the visual appearance of some reconstructions without altering the reconstruction procedure or recovering additional information from the observed gradients.

\section{Additional Qualitative Results} \label{additional results}
\begin{figure}[tb] 
    \centering
    \begin{subfigure}{1\textwidth}
        \centering
        \includegraphics[width=1\linewidth]{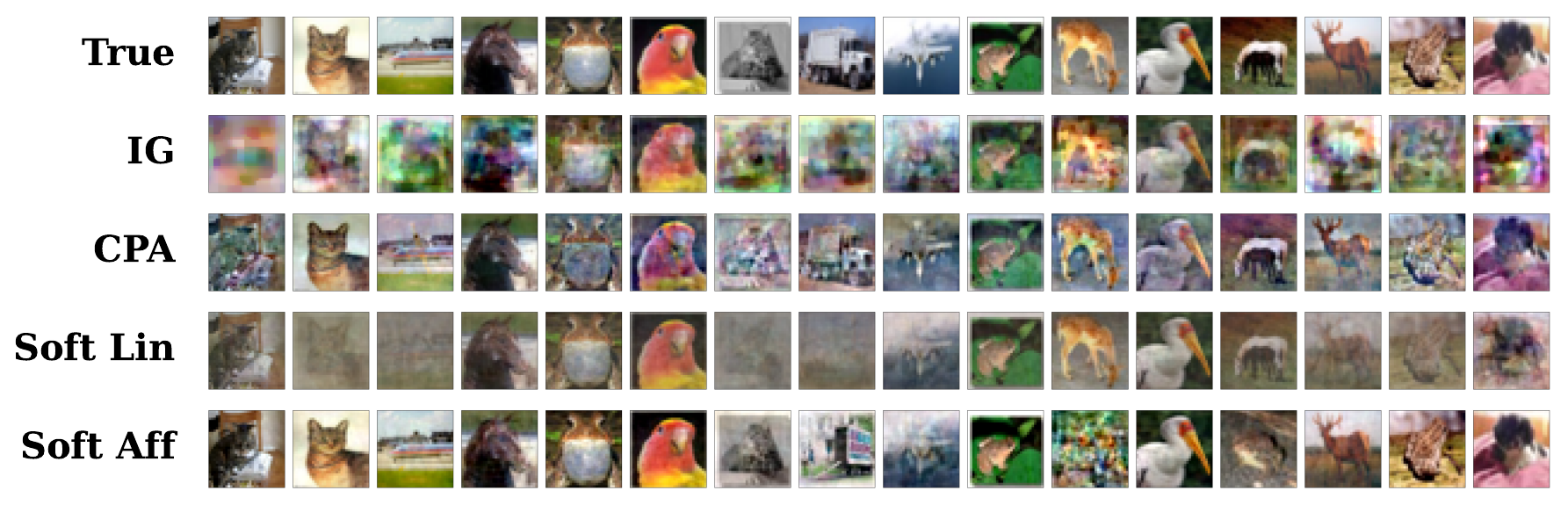}
        \caption{}
    \end{subfigure}
    
    \begin{subfigure}{1\textwidth} 
        \centering
        \includegraphics[width=1\linewidth]{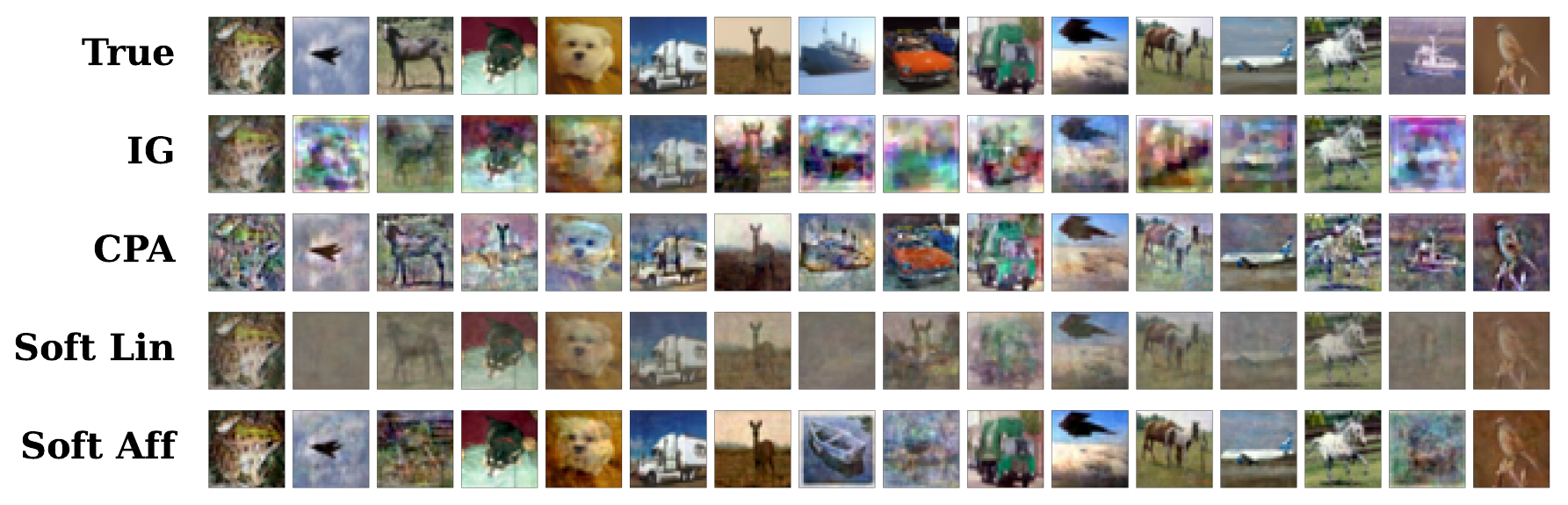}
        \caption{}
    \end{subfigure}

    \begin{subfigure}{1\textwidth} 
        \centering
        \includegraphics[width=1\linewidth]{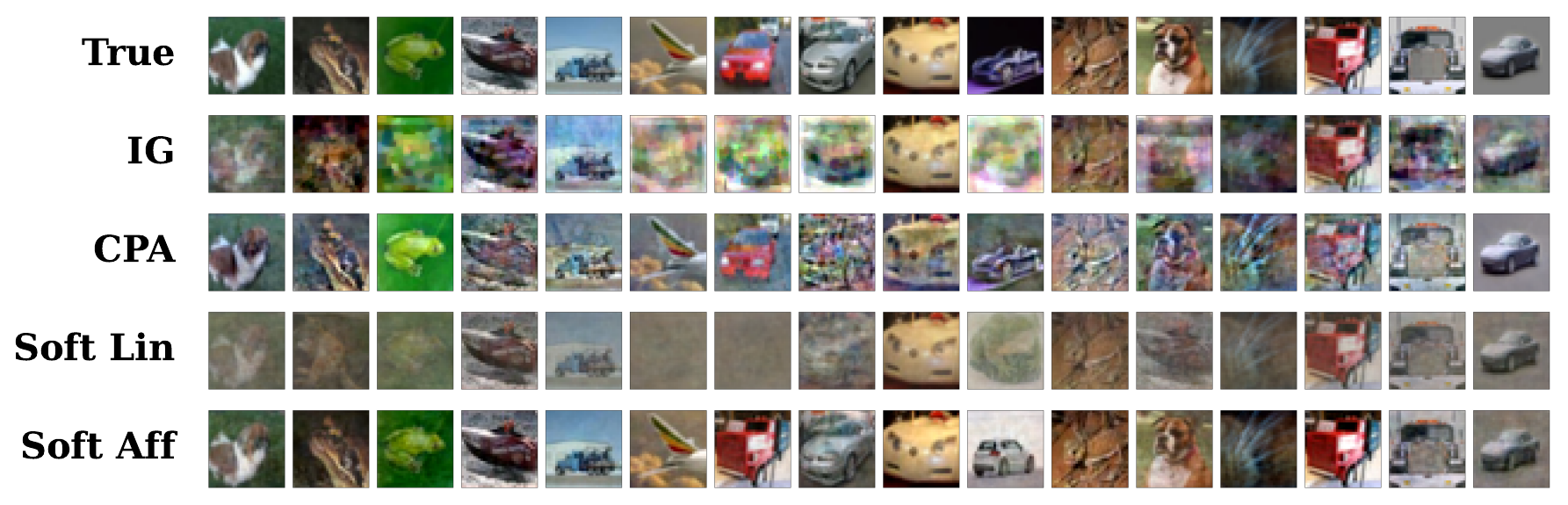}
        \caption{}
    \end{subfigure}
    \caption{Comparison of CIFAR-10 reconstructions for FC-3 with ReLU in the noiseless setting with known batch size $B=64$. A subset of the ground-truth batch is shown alongside the corresponding matched reconstructions from IG, CPA, Soft Linear and Soft Affine. Panels (a)--(c) show examples from 3 different batches.}
    \label{fig:cifar-b64}
\end{figure}

\begin{figure}[tb] 
    \centering
    \begin{subfigure}{1\textwidth}
        \centering
        \includegraphics[width=1\linewidth]{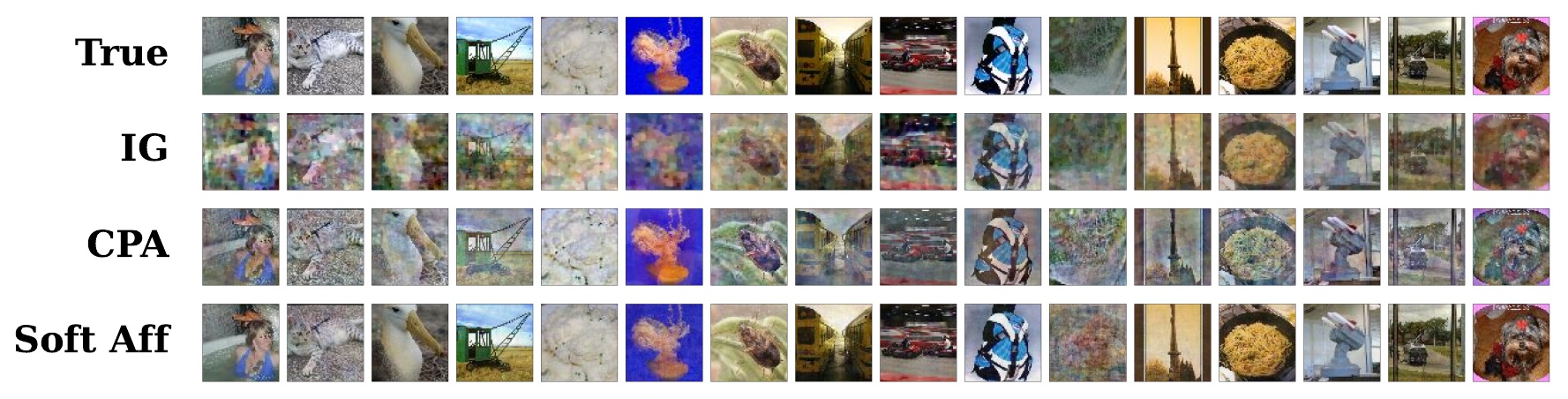}
        \caption{}
    \end{subfigure}
    
    \begin{subfigure}{1\textwidth} 
        \centering
        \includegraphics[width=1\linewidth]{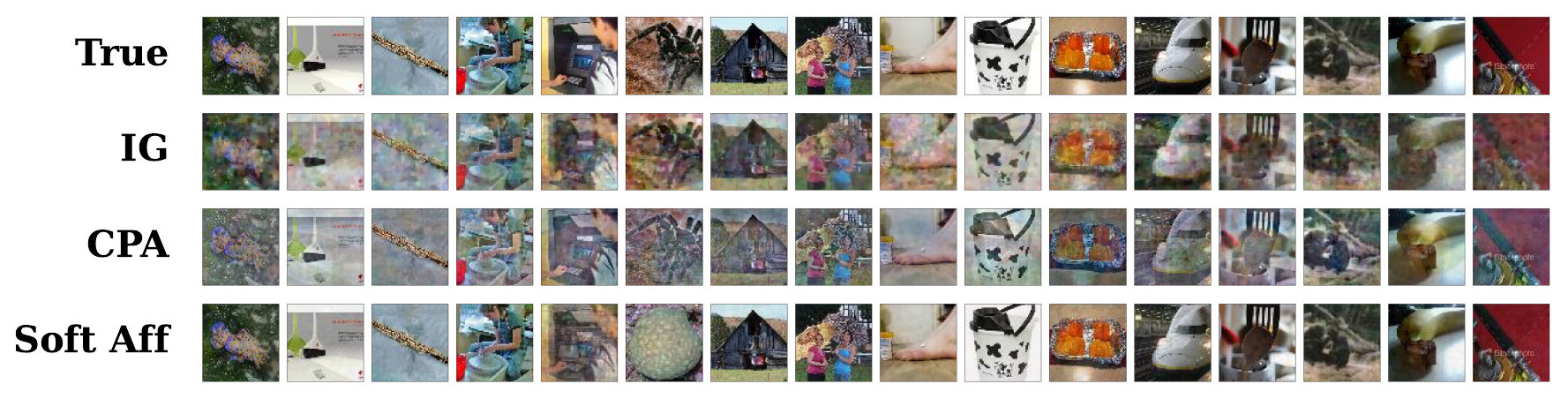}
        \caption{}
    \end{subfigure}

    \begin{subfigure}{1\textwidth} 
        \centering
        \includegraphics[width=1\linewidth]{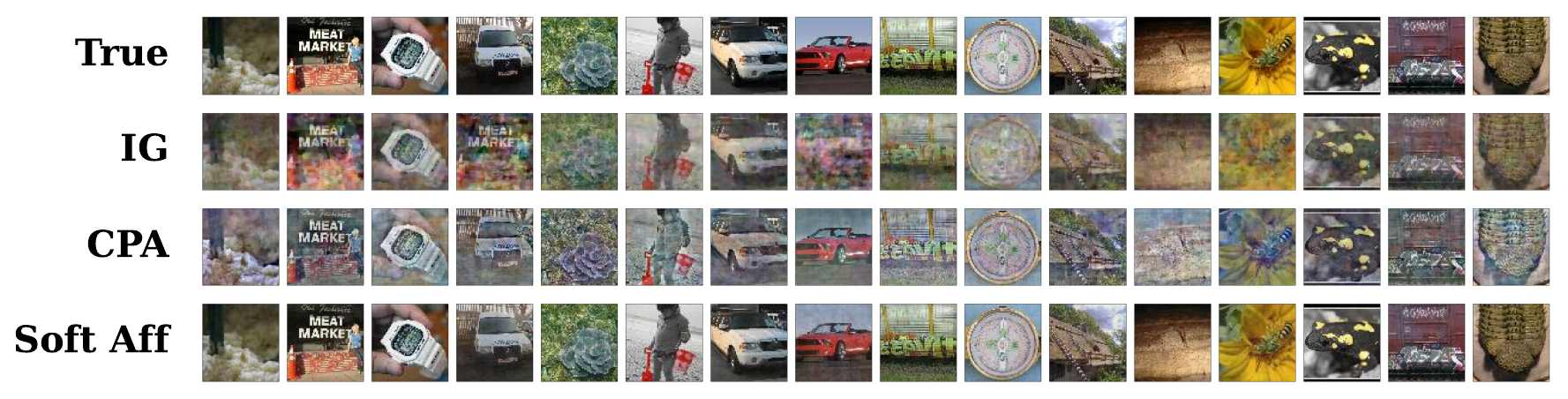}
        \caption{}
    \end{subfigure}

    \begin{subfigure}{1\textwidth} 
        \centering
        \includegraphics[width=1\linewidth]{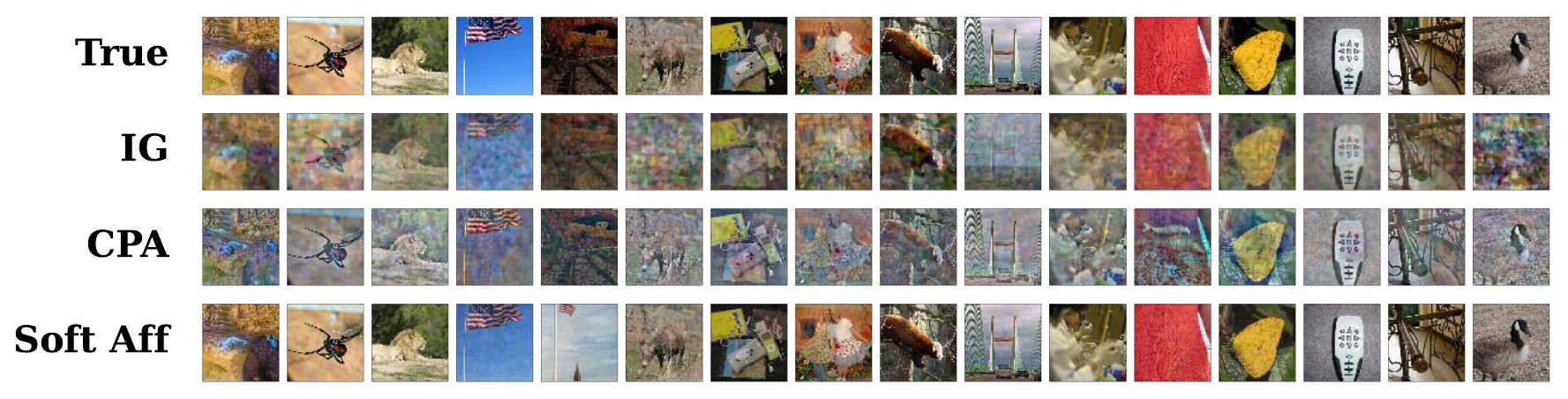}
        \caption{}
    \end{subfigure}
    \caption{Comparison of Tiny-ImageNet reconstructions for FC-3 with LeakyReLU in the noiseless setting with known batch size $B=128$. A subset of the ground-truth batch is shown alongside the corresponding matched reconstructions from IG, CPA and Soft Affine. Panels (a)--(d) show examples from 4 different batches.}
    \label{fig:additional-tiny-imagenet-b128}
\end{figure}

\begin{figure}[tb] 
    \centering
    \begin{subfigure}{1\textwidth}
        \centering
        \includegraphics[width=1\linewidth]{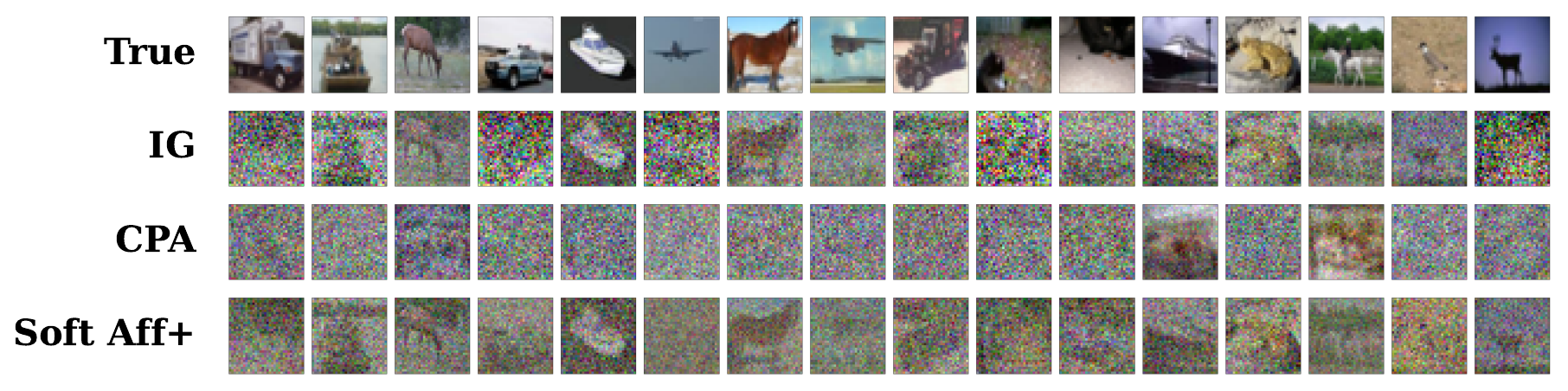}
        \caption{$0$ dB}
    \end{subfigure}
    
    \begin{subfigure}{1\textwidth} 
        \centering
        \includegraphics[width=1\linewidth]{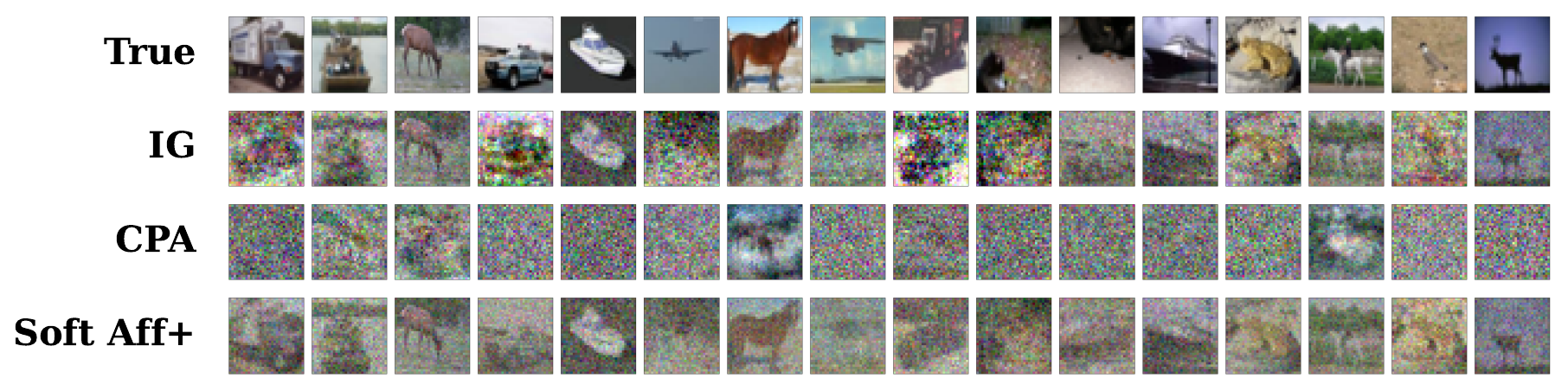}
        \caption{$5$ dB}
    \end{subfigure}

    \begin{subfigure}{1\textwidth} 
        \centering
        \includegraphics[width=1\linewidth]{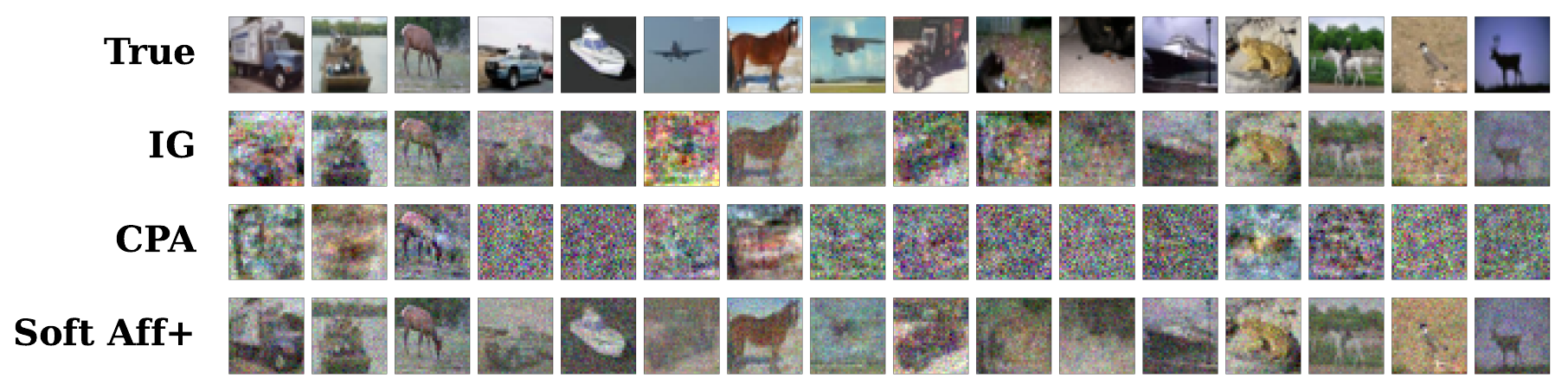}
        \caption{$10$ dB}
    \end{subfigure}

    \begin{subfigure}{1\textwidth} 
        \centering
        \includegraphics[width=1\linewidth]{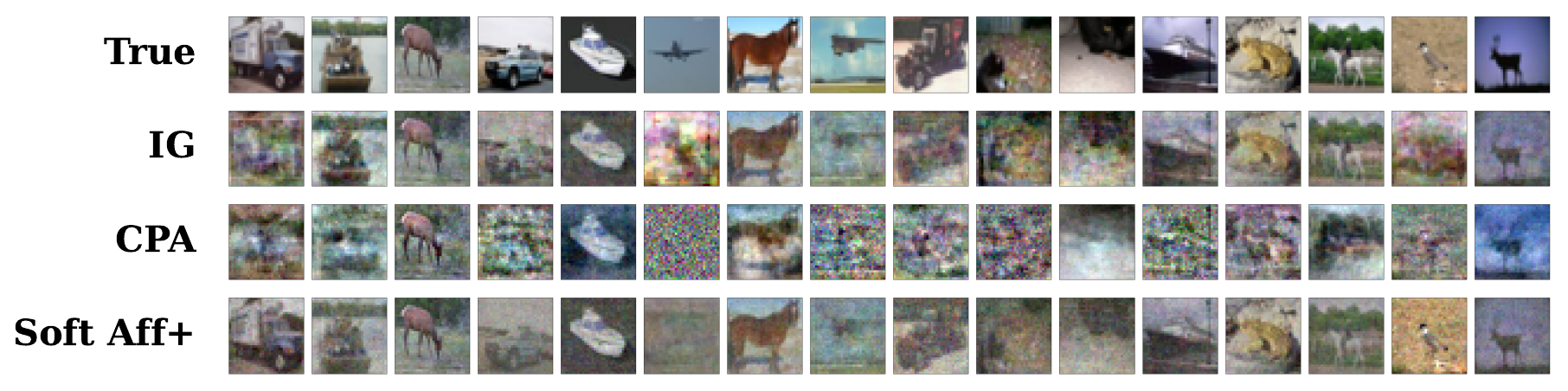}
        \caption{$15$ dB}
    \end{subfigure}
    \caption{Comparison of CIFAR-10 reconstructions for FC-3 with ReLU under gradient perturbations at target SNR levels of (a) $0$, (b) $5$, (c) $10$ and (d) $15$ dB. The same 16 ground-truth images from a batch of $B=32$ are shown across all panels, together with their matched reconstructions from IG, CPA, and Soft Affine. The $+$ suffix denotes percentile-based contrast stretching applied after reconstruction.}
    \label{fig:cifar-noisy-reconstructions}
\end{figure}

\begin{figure}[tb] 
    \centering
    \begin{subfigure}{1\textwidth}
        \centering
        \includegraphics[width=1\linewidth]{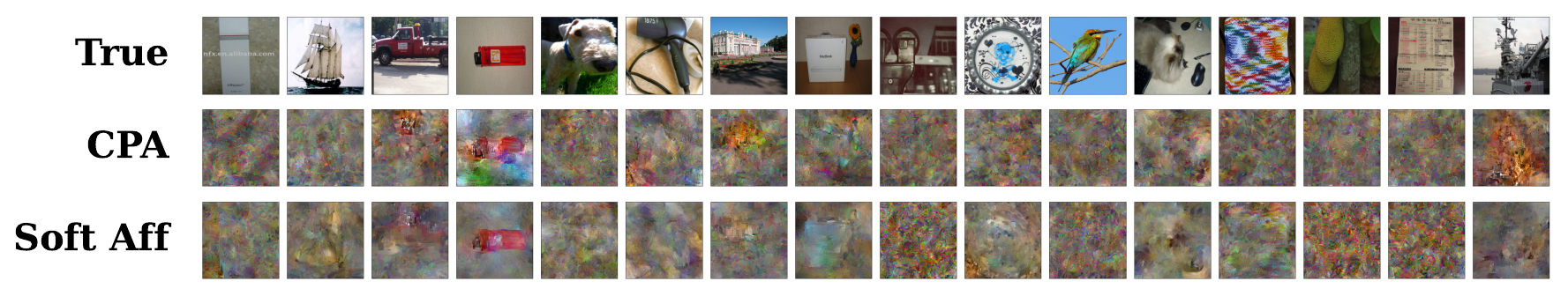}
        \caption{$-15$ dB}
    \end{subfigure}
    
    \begin{subfigure}{1\textwidth} 
        \centering
        \includegraphics[width=1\linewidth]{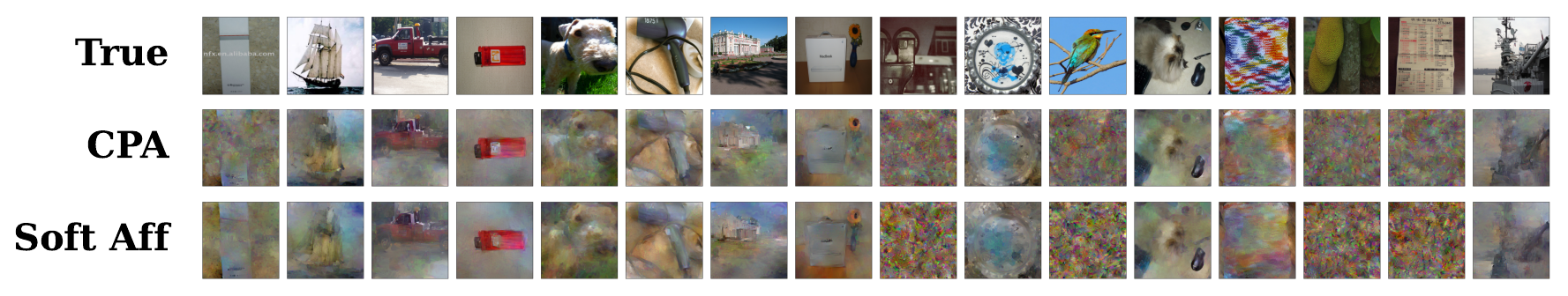}
        \caption{$-10$ dB}
    \end{subfigure}

    \begin{subfigure}{1\textwidth} 
        \centering
        \includegraphics[width=1\linewidth]{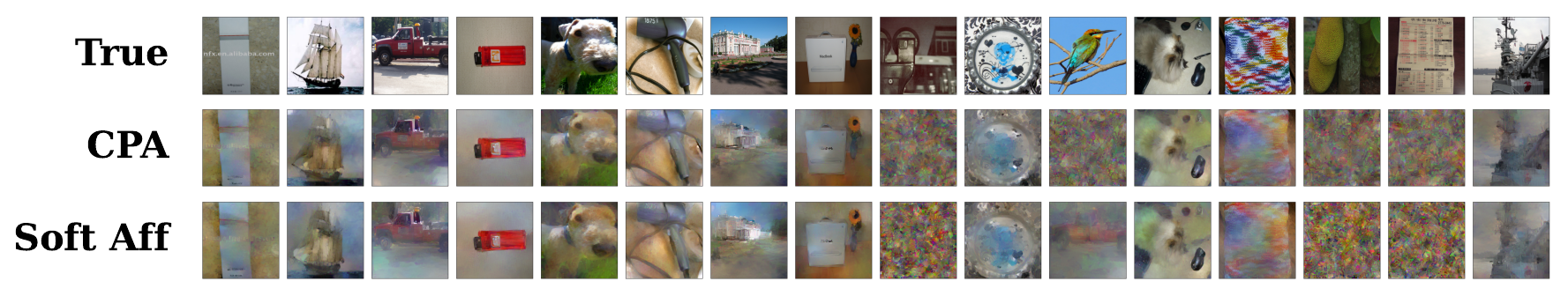}
        \caption{$-5$ dB}
    \end{subfigure}

    \begin{subfigure}{1\textwidth} 
        \centering
        \includegraphics[width=1\linewidth]{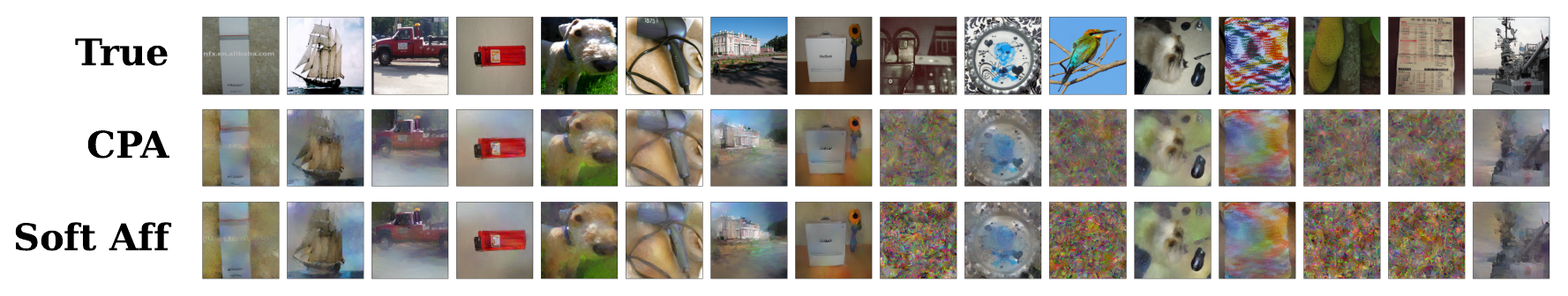}
        \caption{$0$ dB}
    \end{subfigure}
    \caption{Comparison of ImageNet reconstructions for VGG-16 under gradient perturbations at target SNR levels of (a) $-15$, (b) $-10$, (c) $-5$ and (d) $0$ dB. The same 16 ground-truth images from a batch of $B=32$ are shown across all panels, together with their matched reconstructions after feature inversion of embeddings recovered using CPA and Soft Affine. The feature inversion stage also incorporates gradient matching against the observed perturbed gradients.}
    \label{fig:imagenet-noisy-reconstructions}
\end{figure}

This appendix provides additional qualitative examples complementing the quantitative results in Section \ref{section 5}. Figures \ref{fig:cifar-b64} and \ref{fig:additional-tiny-imagenet-b128} show noiseless reconstructions on CIFAR-10 and Tiny-ImageNet, respectively, with Figure \ref{fig:cifar-b64} also comparing the linear and affine formulations. Figures \ref{fig:cifar-noisy-reconstructions} and \ref{fig:imagenet-noisy-reconstructions} illustrate the effect of gradient perturbations on CIFAR-10 and ImageNet reconstruction, keeping the displayed ground-truth images fixed across SNR levels. These examples illustrate differences between methods and the variation in reconstruction quality across individual images.

